\documentclass[sigconf,10pt]{acmart}

\usepackage{amsmath}

\usepackage{algorithmic}
\usepackage[linesnumbered, ruled,vlined]{algorithm2e}
\usepackage{caption}
\usepackage{subfigure}

\AtBeginDocument{%
  \providecommand\BibTeX{{%
    \normalfont B\kern-0.5em{\scshape i\kern-0.25em b}\kern-0.8em\TeX}}}

\setcopyright{acmcopyright}
\copyrightyear{2021}
\acmYear{2021}
\acmDOI{10.1145/1122445.1122456}

\acmConference[xxxxx 'xx]{xxxxx 'xx: xx xxth
xxxxx xxxxx xxxx xx}{xx xx--xx, xx}{xxxx, xxxx}
\acmBooktitle{xxxx 'xx: xxx xxth
xxxx xxxx xxx xxxxx,
  xxxx xx--xx, xxxxx, xx xx, xxxx}
\acmPrice{15.00}
\acmISBN{978-1-4503-XXXX-X/18/06}

\begin{document}

%%
%% The "title" command has an optional parameter,
%% allowing the author to define a "short title" to be used in page headers.
\title{REACT: Distributed Mobile Microservice Execution Enabled by Efficient Inter-Process Communication}

%%
%% The "author" command and its associated commands are used to define
%% the authors and their affiliations.
%% Of note is the shared affiliation of the first two authors, and the
%% "authornote" and "authornotemark" commands
%% used to denote shared contribution to the research.

% TODO: uncomment after review
\author{Chathura Sarathchandra}
\orcid{0000-0002-0266-0446}
\affiliation{%
  \institution{InterDigital Europe}
  \city{London}
  \country{United Kingdom}
}\email{chathura.sarathchandra@interdigital.com}

%\author{xxxx xxxx xxxxx xxxx}
%\affiliation{%
%  \institution{xxxxxxx}
%  \city{xxxx}
%  \country{xxxxxxx}
%}\email{xxxxxx.xxxxx}

%\author{Dirk Trossen}
%\authornote{Part of his work was
%done at the InterDigital Europe, London, United Kingdom}
%\affiliation{%
%  \institution{InterDigital Europe}
%  \city{London}
%  \country{United Kingdom}
%}\email{dirk.trossen@interdigital.com}

%%
%% By default, the full list of authors will be used in the page
%% headers. Often, this list is too long, and will overlap
%% other information printed in the page headers. This command allows
%% the author to define a more concise list
%% of authors' names for this purpose.

%TODO: Correct 
\renewcommand{\shortauthors}{Chathura, Sarathchandra}
%\renewcommand{\shortauthors}{xxxx, xxxx}

%%
%% The abstract is a short summary of the work to be presented in the
%% article.
\begin{abstract}
The increased mobile connectivity, the range and number of services available in various computing environments in the network, demand mobile applications to be highly dynamic to be able to efficiently incorporate those services into applications, along with other local capabilities on mobile devices. However, the monolithic structure and mostly static configuration of mobile application components today limit application's ability to dynamically manage internal components, to be able to adapt to the user and the environment, and utilize various services in the network for improving the application experience.

In this paper, we present REACT, a new Android-based framework that enables apps to be developed as a collection of loosely coupled microservices (MS). It allows individual distribution, dynamic management and offloading of MS to be executed by services in the network, based on contextual changes. REACT aims to provide i) a framework as an Android Library for creating MS-based apps that adapt to contextual changes ii) a unified HTTP-based communication mechanism, using Android Inter-Process Communication (IPC) for transporting requests between locally running MS, while allowing flexible and transparent switching between network and IPC requests, when offloading. We evaluate REACT by implementing a video streaming app that dynamically offloads MS to web services in the network, adapting to contextual changes. The evaluation shows the adaptability to contextual changes and reductions in power consumption when offloading, while our communication mechanism overcomes performance limitations of Android IPC by enabling efficient transferring of large payloads between mobile MS.

%% For achieving this, REACT overcomes the memory and performance issues of Android IPC with transferring large payloads, by providing an efficient application-layer memory space that is integrated into Android IPC communication.
\end{abstract}

%%
%% The code below is generated by the tool at http://dl.acm.org/ccs.cfm.
%% Please copy and paste the code instead of the example below.
%%


%%
%% Keywords. The author(s) should pick words that accurately describe
%% the work being presented. Separate the keywords with commas.

%% add page numbers
\settopmatter{printfolios=true}

%%
%% This command processes the author and affiliation and title
%% information and builds the first part of the formatted document.
\maketitle

\section{Introduction}
The mobile device has become the primary device of choice for most users, providing access to most applications and services at their fingertips, from multimedia to personal finance applications. This is enabled by their continuously increasing computing capabilities, high bandwidth and low latency network technologies. According to Cisco’s Trend Report \cite{CiscoIR20182023}, by year 2023 global mobile devices will grow to 13.1 billion from 8.8 billion in 2018. Moreover, advancements in various software (e.g., Machine Learning algorithms) and human-computer interaction (e.g., Augmented Reality) technologies have helped popularize mobile applications and services providing enhanced user experiences and functionality. Thus, gaming, social media and business mobile apps are predicted to be the most popular out of all 299.1 billion downloaded globally by 2023.

Resources and functionalities required for those applications, in turn, increase the demand on resources on mobile devices that they run on, which are inherently resource-constrained. Therefore, the lack of resources in mobile devices limits the user experience and the type of applications and services that can be offered.

% Existing mobile offloading solutions and having only one server counterpart
Mobile cloud computing \cite{doi:10.1002/wcm.1203} bridges the gap between the growing resource demands of mobile applications, and limited resource availability of mobile devices, through offloading resource-intensive functions to resource-rich application execution environments (e.g., edge, cloud environments) in the network or to other devices. This brings the resource elasticity of cloud computing technology to the much rigid mobile devices. However, existing mobile function offloading frameworks rely on the availability of pre-deployed framework-specific server counterparts in the network \cite{10.1145/1966445.1966473, 10.1145/1814433.1814441, 6195845} to be able to offload computing functions, limiting the overall offloading opportunity. These server counterparts receive and execute offloaded tasks, by following framework-specific execution models and protocols. Partitioning of application components are performed at code level (e.g., through code annotations), maintaining the monolithic code structure, which limits the flexibility and dynamicity in managing independent components. Thus, the application components themselves are monolithically packaged and statically configured for using known server offloading counterparts. This limits the ability of the applications to dynamically use and adapt to newly available services (such as microservice clouds \cite{8449768}) and new execution environments, respectively.

% What influences offloading performance -  and importance of adaptability to contextual changes
However, offloading is not always beneficial and may lead to negative performance gains \cite{5445167}. For example, in cases where it requires large amount of data to be transferred between the mobile device and the cloud when offloading, it may lead to higher execution times as well as increased power consumption. However, under certain conditions the user may prefer to offload functions for obtaining a better functionality and a better user experience, at the expense of increased energy consumption. Therefore, dynamically adapting to various contextual changes \cite{7362223} when offloading is crucial, while decisions on when to offload, what to offload (which functions) and where to offload (execution environments and other mobile and Internet Of Things devices) have a direct impact on the overall performance of the mobile device.

% service availability as a primary factor needed to improve offloading opportunity
The availability of services in the network is one of the primary factors that influence mobile function offloading decisions. If there are more services that can be utilized for offloading, the more opportunities and options (e.g., multiple service providers offering same service with varying qualities) the applications have when deciding when and where to offload \cite{SARATHCHANDRAMAGURAWALAGE201422}. Moreover, increased bandwidth, lowered latency and newly introduced edge computing resources in emerging new communication technologies such as 5G \cite{7931566}, do not only increase the efficiency with which those services could be accessed and executed, but also increase offloading opportunities by allowing application and service providers to deploy application and service instances both in the edge and in the cloud.

% But static binding is an issue with mobile function offloading frameworks
As a result of the increased ubiquity in services, resource provisioning extends beyond public cloud and the mobile functions may be offloaded to services in the edge as well as to services offered by other mobile/IoT devices. However, as discussed above, monolithic design and statically binding an application to one specific resource or to a specific application server counterpart, offered by the same application provider, limits offloading opportunities that a device may have for improving performance or augmenting functionality. For example, an application provider may deploy instances of the same service at multiple mobile, edge and cloud resources, and a device that is bound only to a specific resource or a server instance in the cloud may not be able to utilize an instance that is closer to the edge or offered by another IoT device within user's proximity. Moreover, with the increased range of web services provided by the application providers as well as third-party service providers that are available in networks (e.g., varying cloud video encoding/decoding services), allowing applications to dynamically utilize third-party services for offloading mobile functions will further increase offloading opportunities and reduce costs in service provisioning. Therefore, it is imperative that applications can dynamically utilize those third-party services, as opposed to statically binding to a specific server instance in the network provided by the corresponding application provider.

%microservices is the answer
Microservices architecture \cite{8354433} has become an increasingly popular application structure style and microservice clouds have been used for providing services for various applications \cite{8449768}. It provides applications with the flexibility to adapt to the underlying technological changes and, the ability to manage resources and control different application functions (i.e., microservices) independently. In this architecture, an application is structured as a set of loosely coupled and independently manageable services that communicate over a lightweight (and often technology-agnostic) protocol (such as HTTP \cite{fielding1999hypertext}). The fine-grained control of constituent components enabled by the microservices architecture, allows the internal structure of applications to be dynamically changed at runtime. The use of a common communication interface (an application-layer protocol such as HTTP) across all microservices enables dynamic binding to offloading counterparts, i.e., offloading to other (micro)services over the network. This enables dynamic mobile function offloading and utilization of suitable microservices deployed by the application provider or other third-party service providers. Moreover, with the increased availability, range and number of services (and microservices), mobile offloading frameworks using microservices architecture can significantly increase offloading opportunity and flexibility in choosing the best services to be used towards improving application functionality and performance. Therefore, the flexibility and adaptability (to contextual changes) of mobile applications can be significantly improved through combining the high flexibility of microservices architecture and the dynamicity of mobile function offloading frameworks.

This paper presents REACT (mic\textbf{R}oservice \textbf{E}xecution en\textbf{A}bled by effi\textbf{C}ient in\textbf{T}er-process communication) \footnote{An early prototype was demonstrated at a major leading congress in 2019, and the complete system was to be demonstrated in 2020}, a framework (made available as an open-source Android Library \footnote{The Link to open-source code of REACT Android Library - https://github.com/chathura77/REACT}) which allows Android mobile applications to be developed as a collection of loosely coupled microservices that can be independently managed. Using the framework, the Android application developers can incorporate into their applications the flexibility provided by the microservices architecture and the adaptability of applications to contextual changes of mobile function offloading frameworks (by offloading functions to improve functionality and/or performance). REACT can be integrated into an application by simply including the provided Android Library and using the provided easy to use Application Programming Interface (API). Using REACT, applications can dynamically (based on contextual changes) utilize web services for offloading various mobile Application Functions (AF)/microservices, towards improving performance and functionality of mobile devices, i.e., the presented API allows the developer to modularize an application into constituent microservices, which then in turn can be offloaded independently to be executed on corresponding web services. A new mobile microservice communication mechanism is introduced, which uses the HTTP protocol for communicating between microservices and services in the network, while enabling flexible and dynamic switching between requests to microservices on mobile device and web services in the network, when offloading at runtime, transparently to communicating microservices. Our solution uses Android Inter-Process Communication (IPC) mechanism for transferring messages between communicating microservices within the mobile device. Thus, REACT tackles critical performance limitations incurred by Android IPC when transferring large payloads \cite{10.1145/2436196.2436200}. In the rest of this paper we use 'Application Functions' (AF) and 'Microservices' interchangeably when referring to application components modularized using REACT.

Our main contributions are;
\begin{itemize}
\item An android application framework which enables modularization of app components into microservices for context-aware independent management and offloading of mobile microservices, at runtime.
\item A HTTP-based mobile microservice communication mechanism, enabled by Android IPC-based efficient local inter-microservice communication, and flexible switching between local and network communication for dynamic microservice offloading.
\end{itemize}

\section{Related Work}
There have been several extensive studies on mobile computation offloading frameworks in the past years \cite{6365155}. These approaches intend to augment capabilities of resource-constrained mobile devices by dynamically offloading computationally intensive tasks to resource-rich destinations. Existing frameworks use services from surrounding computing devices \cite{10.1145/3300061.3345443, 10.1145/3307334.3326096, 10.1145/1839294.1839332} or remote cloud machines \cite{10.1145/1966445.1966473, 10.1145/1814433.1814441, 6882212, 6898822, OSULLIVAN201520, 6195845}, where specific protocols and offloading server counterparts are deployed at the offloading destinations, and does not incorporate the microservices architecture i.e., offloading framework specific server counterparts that support their offloading protocol needs to be installed at either surrounding computing devices or remote cloud machines, prior to execution of the application. While work presented in this paper is inspired by insights provided in previous work, REACT focuses on creating flexible mobile applications based on the microservice architecture supporting seamless offloading of local application functions to web services.

MAUI \cite{10.1145/1814433.1814441} framework allows the developer to provide an initial partition of the application through remotable annotations, indicating methods and classes that can be offloaded to the MAUI server. The main aim of MAUI is to optimize energy consumption by offloading suitable computing intensive application tasks. MAUI runtime on smartphone communicates with the MAUI runtime running on the MAUI server over Remote Procedure Calls when offloading. Likewise, ThinkAir \cite{6195845} provides method-level task offloading, based on code annotations provided by the developer at design time. ThinkAir code generator then generates remotable method wrappers and other utility functions required for offloading and remote execution of application tasks. Moreover, ThinkAir optimizes application execution by allowing parallel execution of offloaded tasks, speeding up the execution. Both MAUI and ThinkAir gather various hardware, software and network related information through profiling to be used as inputs when making offloading decisions.

CloneCloud \cite{10.1145/1966445.1966473} provides runtime partitioning of applications at thread level based on a combination of static analysis and dynamic profiling techniques. The system enables partitioning of unmodified applications, i.e., partition without developer’s help. The offloaded tasks are executed on a pre-deployed cloned Virtual Machine (VM) in cloud with higher resource capacity. The application-layer VM-based offloading mechanism is used, where states offloading threads are encapsulated in VMs and them migrated to remote cloud for execution. The execution runtime includes a per-process migrator thread and per-node node manager, for managing thread states (i.e., suspending, packaging, resuming and merging), and for managing nodes (i.e., device-to-clone communication, clone image synchronization and provisioning), respectively. Moreover, the CloneCloud system offloads towards optimizing execution time and energy consumption. Both the local device and the clone are devised with corresponding manager, migrator and profiler counterparts, alongside the application.

Augmenting computational capabilities of mobile devices by offloading tasks at the service level granularity has been studied in \cite{6882212, 7037763}, where the tasks are offloaded to a remote cloud server that provides computing task execution services as RESTful services. Elgazzar K, et. al \cite{6882212} presents a task offloading framework which takes the location of the data required for executing offloaded tasks (in a scenario where data is provided by a third-party provider) in addition to the capabilities of the cloud service provider, when making offloading decisions. RMCC \cite{7037763} framework employs a RESTful API for offloading tasks and takes device energy consumption and execution time when making offloading decisions.

Conventionally, computation task offloading decisions are made for reducing the response time and energy consumption or for exploiting a balance between the two. The seminal work of Kumar et al \cite{5445167} provided an analytical mode with which one may determine whether offloading tasks can save energy, taking computing resource requirements and instantaneous network condition into consideration. Such decisions may be performed either using rule/policy-based approaches \cite{10.1145/1814433.1814441, 6195845, 10.1145/1966445.1966473} or based on learning techniques \cite{6809335}. In general, most offloading decision-making strategies consider a combination of factors such as energy consumption, task complexity, network condition, computing resource availability \cite{SARATHCHANDRAMAGURAWALAGE201422} and utilization. 

\section{Design Goals \& Architecture}

The ultimate goal of REACT is to improve performance and augmentation of capabilities of mobile devices through utilization of available web services and computing resources in the connected environments. In this section, we present the design objectives of REACT, followed by a high-level overview of REACT’s components.

\begin{enumerate}

\item \emph{Microservices for mobile applications}:
Enable mobile applications to be developed as a collection of loosely coupled app functions (i.e., microservices), that communicate using the HTTP protocol \cite{fielding1999hypertext}. allowing them to be independently managed. 

\item \emph{Dynamic adaptability to contextual changes}:
Key characteristics of mobile computing environments are their rapid change and volatility. Therefore, any solution that operate in such an environment must adapt to the changes and account for volatility, towards ensuring the requirements of the application execution. REACT enables applications to adapt to contextual changes through allowing functions to be dynamically offloaded based on contextual information gathered from the mobile device.

\item \emph{Compatibility with web services}: Provide the ability to utilize and incorporate web services in the network as part of the mobile application. Allow mobile devices to dynamically utilize web services over HTTP for mobile function offloading, enabling dynamic binding to offloading server counterparts/web services.

%\item \emph{Ease of use for developers}: 
%Application developers are required to identify and define application tasks at the design phase, which are suitable for offloading. Although, the underlying mechanisms of mobile task offloading frameworks are part of the application, the application developers ideally should not be burdened with such detail and require only to focus on the design and development of the application itself. REACT provides an easy to use API which can be used by the developer by including the provided library with the application without needing prior knowledge of the underlying mechanisms. The provided library can easily be included in the application just like any other library and follow the simple guidelines provided for encapsulating application tasks into modular application functions which can be offloaded, which even a novice developer with entry level programming knowledge is able to follow.

\item \emph{Flexible and efficient communication mechanism}:
Introduce an efficient and flexible communication mechanism unifying inter-AF communication both within device and with services in the network. Use IPC mechanisms provided by Android OS for local AF communication, while also allowing the AF message transport technology (i.e., TCP or Android IPC) to be dynamically switched, transparently to the communicating AFs, depending on their locality.

\end{enumerate}

\begin{figure}
  \centering
  \includegraphics[width=\linewidth]{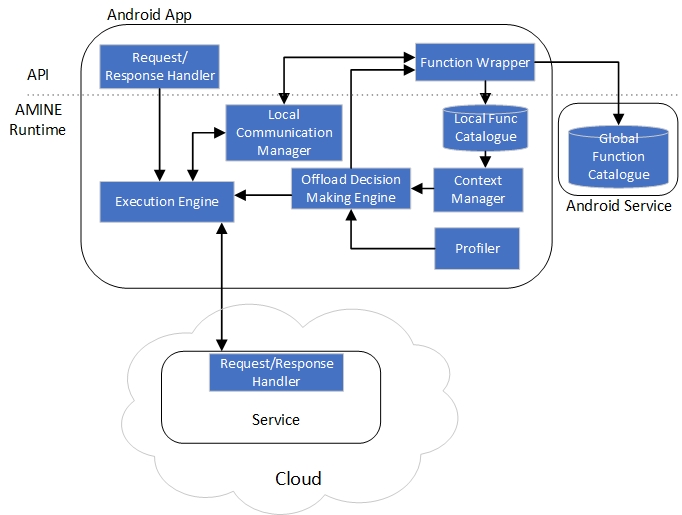}
  \caption{Overview of the REACT framework.}
  \Description{}
  \label{fig:architecture}
\end{figure}

As depicted in Figure \ref{fig:architecture}, the REACT framework has two major aspects, the API (Section \ref{sec:api}) and the REACT Runtime (Section \ref{sec:runtime}). The new IPC-based communication mechanism is introduced in Local Communication Manager (Section \ref{sec:lcm}). The following sections present details of each of those components.

\begin{figure}
  \centering
  \includegraphics[scale=0.85]{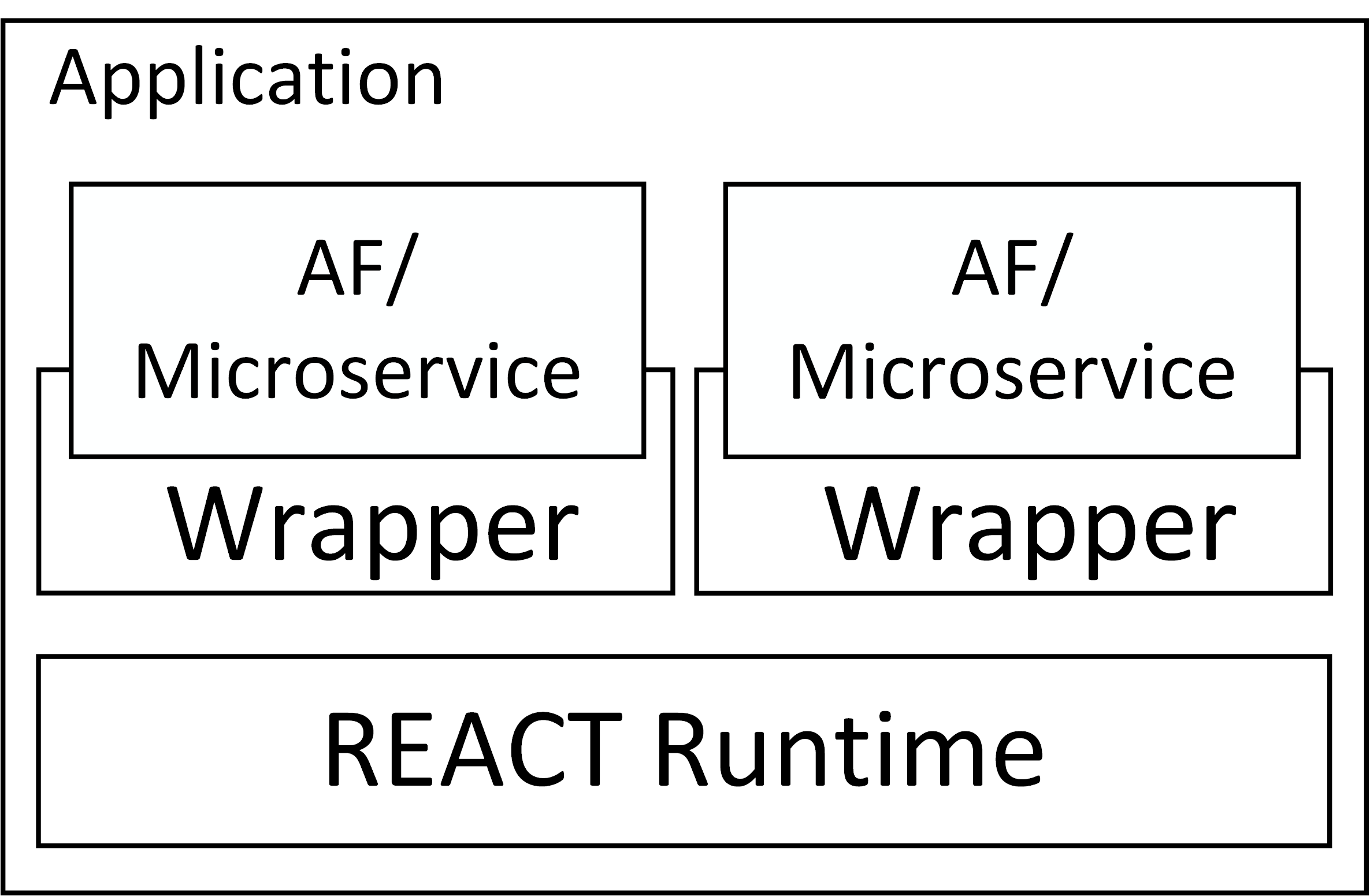}
  \caption{REACT API Overview}
  \Description{}
  \label{fig:runtime}
\end{figure}

\section{Programming Interface} \label{sec:api} % DONE: add that AF and microservices terms are used interchangeably

The REACT framework provides the developer with an API which can be used for defining various application tasks as independent Application Functions (AF) that can be automatically offloaded to web services running on execution environments in the network (e.g., edge, cloud). Thus, the REACT API enables the creation of interoperable mobile AFs that can be executed and dynamically managed within the common REACT platform. 

As shown in Figure \ref{fig:runtime}, the REACT library provides a \emph{Function Wrapper} class which can be used for encapsulating and preparing AF procedures and data structures. This allows the developer to define local AFs matching either existing services in the network, or ones that may be deployed later for enabling runtime offloading (e.g., an app developer may develop and deploy remote counterparts to locally running AFs, as web services, which are in turn used for offloading). For example, in a video viewing application the developer may encapsulate the code that process the frames into a "process" AF partition along with its constituent procedures and data structures.

Moreover, using the methods provided with the wrapper class, the developer is able to set AF address, e.g., Fully Qualified Domain Name (FQDN) of the service, and handle AF communications by implementing corresponding virtual request handler  methods, following the request/response model. At the initialization stage, the wrapper class connects with the underlying REACT runtime and automatically configures AF related parameters (e.g., registering the availability of the AF with the Function Catalogue). All other required AF control and management tasks are done automatically by the underlying management components of the REACT framework, interfacing with the corresponding wrapper classes. 

After the initialization phase, AFs become accessible using the provided address, within the device. Any application component (including other AFs) can communicate with the AFs using the "Request" class provided by the framework (or using other provided Request sub-classes which contain methods for parsing and handling responses with specific data types, e.g., \emph{ByteArrayRequest}, \emph{StringRequest}).

\begin{figure}
  \centering
  \includegraphics[scale=0.25]{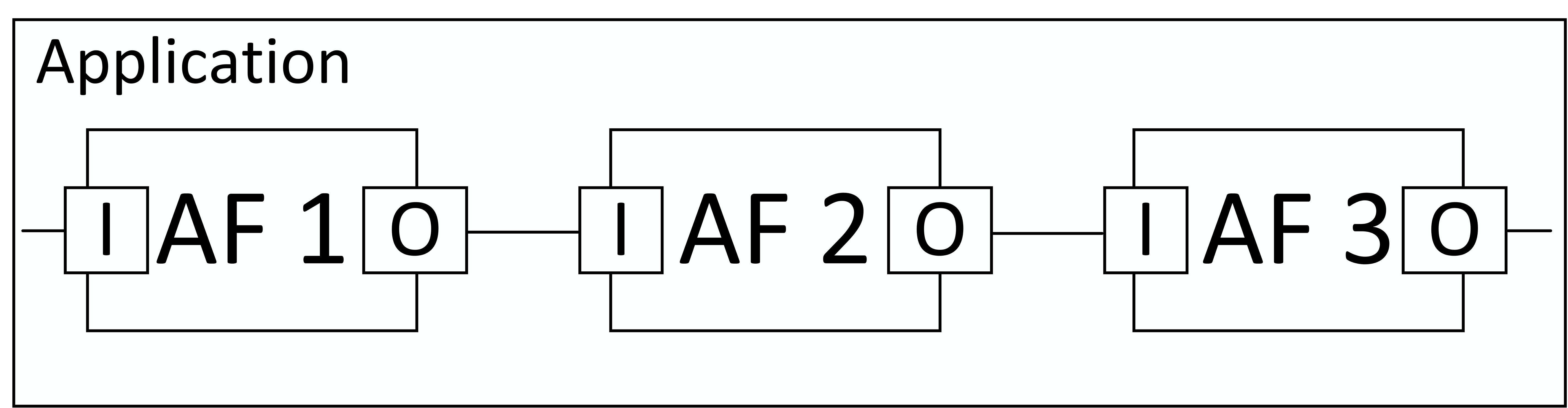}
  \caption{Application following SFC model}
  \Description{}
  \label{fig:chaining}
\end{figure}

The API provides the developer with the flexibility to follow other models of application development and structuring methods. Such examples include Microservices and Service Function Chaining (SFC). Microservices is an architecture that structures an application as a collection of loosely coupled services. REACT API allows one to realize microservice based applications on Android devices, using wrapper classes to construct independently manageable components that communicate with other microservices or services over HTTP. SFC model allows one to structure an application as a collection of sequentially connected services. In the example presented in Figure \ref{fig:chaining}, 'I' indicates the input interface and 'O' indicates the output interface of each AF. Here, the \emph{Function Wrapper} class is used as the input interface for handling incoming requests while the \emph{Request} class is used as the output interface for communicating output data to the next AF in chain.

In all above scenarios, the REACT runtime manages both the execution (dynamically offload suitable AF) as well as the communication between AFs (in-device vs network communication).

\subsection{Function Wrapper}

\begin{figure}
  \centering
  \includegraphics[width=\linewidth]{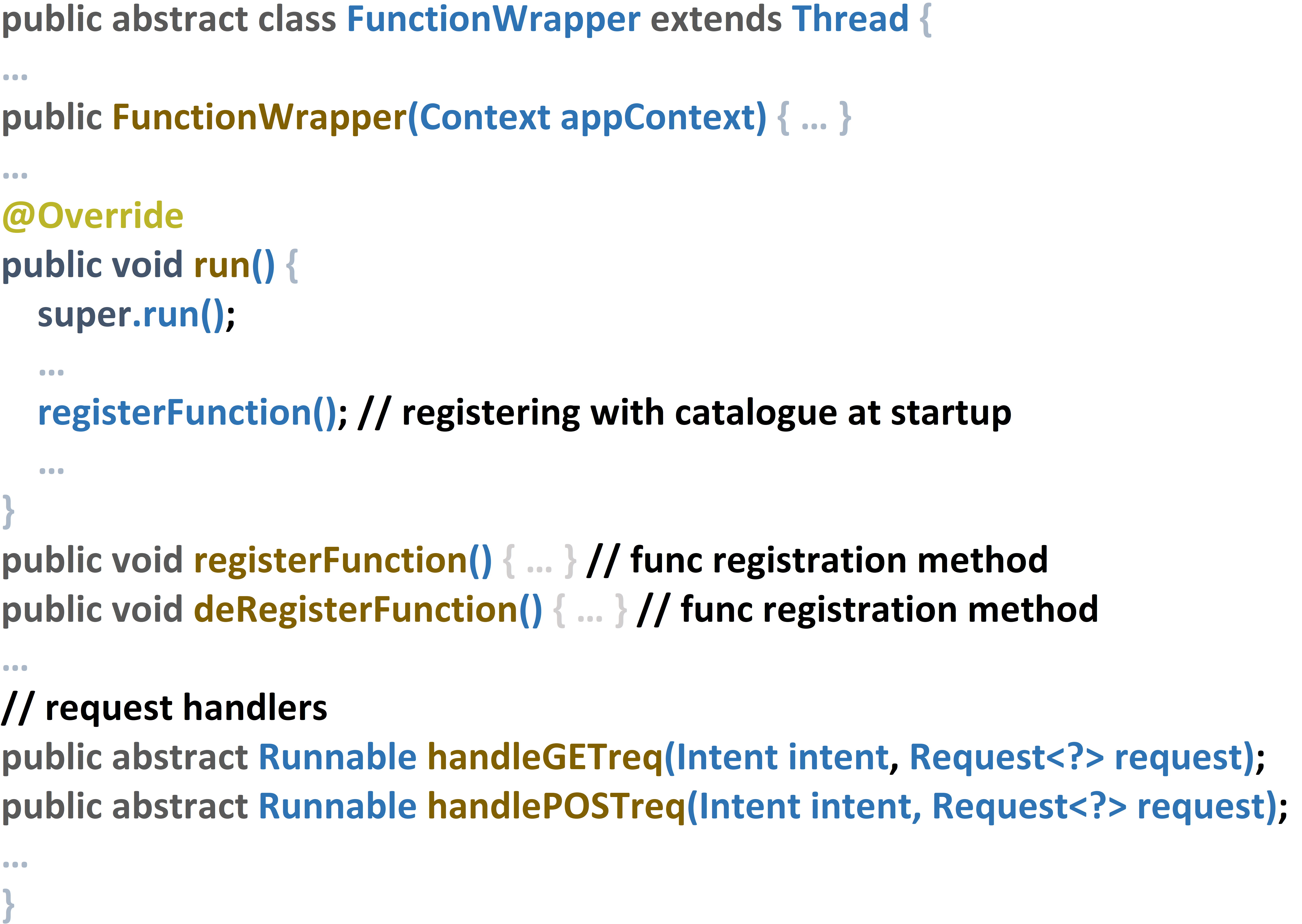}
  \caption{A snippet of code of the Wrapper Class}
  \Description{}
  \label{fig:wrapperclass}
\end{figure}

The Function wrapper (as shown in Figure  \ref{fig:wrapperclass}) class provides the means for any application component in the device to create independent AFs. It provides an interface for AFs to serve incoming requests on a specific URL. The Java abstract methods provided by the interface allow the developer to handle/serve incoming requests, invoking internal AF procedures, in turn linking them to the REACT runtime. The class also includes other helper functions, that are used by the REACT runtime and not used by the developer (methods automatically invoked for registering/deregistering AFs with the Function Catalogues at runtime). The code snippet in Figure \ref{fig:wrapperclass} shows the structure of the interface.

\subsection{Requests}
Requests can be created to request services from AFs or any other web service in the network, in the forms of RESTful (HTTP) or SOAP (Simple Object Access Protocol) requests. In Figure \ref{fig:chaining}, this \emph{Request} class has been used as the output interface, for transferring the output of one AF to the next one. Once, a request is constructed, depending on the locality of the requested AF, i.e., if the AF is executed locally or in the network, the request is indirected to the locally executed AF or sent over to the service in the network, respectively, by the \emph{Execution Engine}.

\section{EXECUTION \& RUNTIME} \label{sec:runtime}

In this section we describe in detail the REACT runtime components that run on the mobile device. These components are included in the REACT framework library provided to the developer and started automatically when the device and the application start. Since conventional web services can be used for offloading computing tasks in the network and REACT does not require any specific components or configurations to be deployed there, only components at the mobile device are discussed. A separate section has been devoted to the \emph{Local Communication Manager} component, in Section \ref{sec:lcm}.

\subsection{Function Distribution and Execution}
Much like the third-party application code that are distributed in the form of libraries today, AFs that use the REACT API (i.e., AFs using the \emph{Wrapper} class) can be distributed to be used independently in other applications that support the REACT framework. We envision that third-party REACT compatible AF providers may also provide compatible matching web services to be distributed in the network, allowing applications to benefit from the dynamic offloading functionality. Developers can include REACT AF in their apps, and access their services through the \emph{Request}s. An AF can be dynamically instantiated within an Android application or an Android service, by simply instantiating its corresponding wrapper class (using Java \emph{new} operator) and invoking \emph{start()} on the AF instance, after setting required parameters (e.g., Android Application Context \cite{androidcontext}).

\subsection{Function Catalogues}

The function catalogue keeps an up-to-date record of all available functions that are available to serve requests within the device. There exist two catalogues, namely, 1) Local (intra app) function catalogue and 2) Global (inter app) function catalogue. The intra-app catalogue keeps record of AFs that are available only within the application, while the inter-app catalogue keeps record of AFs that area available device wide (i.e., implemented as an independent android service, while a AF registration and query interfaces are exposed over Android IPC) All AFs register with the intra-app catalogue as its default behavior, while the application developers can choose to make an AF available to other apps by setting the corresponding 'global catalogue' parameter to 'true' in \emph{registerFunction()} method in the Function Wrapper class. For example, a new functionality may be made available (as an always-on Android Service)  for all applications running locally by a third-party application provider or by Android OS, while still enabling the flexibility of offloading to the network (enabled by REACT framework). Each AF record contains the following tuples.

\begin{itemize}
\item \textbf{Address}:
Each AF is assigned an address. For offloading an AF to be executed by a web services in the network, the address must adhere to the condition that there must be an address mapping logic with which an AF address can be mapped to the URL of the web service in the network. For example, the current implementation uses the reverse-DNS notation of the corresponding web services used for offloading, to name the corresponding AFs, enabling a simple and efficient mapping between local AFs and web services. This mapping logic is realized in \emph{lookup()} function in the \emph{Context Manager}.

\item \textbf{Communication method}:
REACT allows AFs to support multiple communication methods. REACT configures Android Intent broadcast IPC as the default communication method, but a given AF may offer one or more in-device function-to-function communication method for serving requests, using any of existing android IPC mechanisms. REACT allows the developer to specify the supported communication methods when initializing the \emph{Function Wrapper} instance, which are then automatically added into the catalogue. However, it is advisable not to enable more than one communication method per AF for reducing the communication management overhead caused by separate communication types. To overcome shortcomings such as limited bandwidth and inefficient memory usage of Android IPC mechanisms, we introduce our own communication mechanism in Section \ref{sec:applayerheap}, which can used by AFs.
\end{itemize}

\subsection{Context manager}
The \emph{Context Manager} gathers various instantaneous and real-time context information which can be used for optimizing application execution, personalizing the user experience and for dynamically adapting the application to the conditions in the environment. The \emph{Context Manager} is aware of the locally running AFs (received from catalogues as shown in Figure \ref{fig:architecture}) and gathers other information in following categories.

\begin{itemize}
\item \textbf{Location}:
Mobility of the device play an important role in deciding when to offload, e.g., prior knowledge of the web services available within a campus, may allow the users within the campus to offload when in the premises. Most modern mobile devices come with GPS sensors which can be used to obtain user’s geographical coordinates. APIs provided by mobile platforms can be used to access instantaneous GPS information. 

\item \textbf{Device Resources}:
When making offloading decisions, the status of the resources on the device is crucial, (e.g., when offloading for optimizing local resource consumption). Such status information include: CPU utilization, memory utilization, battery charging status, battery level, and storage capacity. For example, battery “plugged” and “unplugged” statuses as well as the current battery level can be used for deciding whether to offload a resource intensive AF when not plugged into a power source.

\item \textbf{Connectivity}:
Mobile devices today come with a range of communication interfaces supporting a wide range of applications. Such interfaces include, Cellular, Bluetooth, WiFi and NFC, each having different characteristics and capabilities. Real-time network connectivity information is imperative to be used for making offloading decisions, For example, an offloading algorithm may consider the instantaneous network throughout and latency information, as they are two parameters that directly affect the performance when offloading. 

\end{itemize}

%\subsection{Profiler}
%This component is responsible for analyzing the characteristics of the AF components (and the overall application) executed on the mobile device. Per each AF, it profiles consumption of resources such as CPU (and the number of threads used), energy, network, memory and storage, as well as interactions with other AFs and other local resources (e.g., android services). These information are collected (offline) when the corresponding AFs are run locally on the device and made available to be used by the \emph{Offload Decision Making Engine}, as an estimation of the hardware and software requirements of each AF.

\subsection{Offload Decision Making Engine}
The \emph{Decision Making Engine} decides on the best execution strategy for AFs, based on information provided by the \emph{Context Manager}. Based on a set of preconfigured policies (or an algorithm) this component decides whether to execute a given AF locally on the mobile device or to offload to be executed by a service in the network. Thus, it invokes the corresponding callback methods ('\emph{offloadFunction}' and \emph{initFunction}) for offloading or executing a AF locally. This, in-turn calls the \emph{Wrapper class} to stop/start the AF, and \emph{Execution Engine} to switch traffic between IPC and network. For example, if a specific AF happens to cost more energy when executed locally than offloaded, the \emph{Offload Decision Making Engine} may decide to offload it when the device is not plugged into power. Similarly, when the device is connected to user's home WiFi network, computing-intensive tasks of a mobile game may be offloaded to a gaming console connected to the same network.

\subsection{Execution Engine}
Requests that are created get added to a First In First Out (FIFO) queue that is managed by the \emph{Execution Engine}. The execution of each request is initiated based on the requested AF’s locality and its chosen communication mechanism, that is provided by the \emph{Offload Decision Making Engine}. This results in either sending the request to a service executed in the network (over the \emph{HttpStack} in REACT that implements HTTP protocol stack), or in-directing it to a locally executing AF (over the \emph{IPCStack}), i.e., if the AF has been decided to be offloaded then the request is sent as a HTTP request to the web service, otherwise, the request is provided to the \emph{Local Communication Manager} for sending it to the locally executing AF using the chosen local function communication mechanism (i.e., any available IPC mechanism). \emph{Local Communication Manager} uses the same communication medium for responses as their corresponding requests. Switching between IPC and network requests are performed on a request-by-request basis, and transparently to the AFs. AFs are agnostic to the underlying communication mechanism used. Therefore, AFs are not aware of weather peering AFs reside locally or in the network and, do not need communication mechanism specific procedures to be implemented within AFs. Moreover, this decoupling of underlying communication and lifecycle management from AF's internal procedures, allows REACT to upgrade existing features without altering AFs themselves (i.e., AF specific internal procedures - and its required to recompile after updating the included REACT library). Further details on how requests are handled by the \emph{Local Communication Manager} when AFs communicate locally are presented in Section \ref{sec:lcm}.

\section{Local Communication Manager}\label{sec:lcm}
When in-directing requests to locally executing AFs, any of the IPC mechanisms offered by the mobile platform may be used. The \emph{Local Communication Manger} (LCM) sends the Request (and receives the Response) over the chosen IPC mechanism, while hiding complexities specific to the local communication mechanisms from the \emph{Execution Engine}. For example, it turns all local AF communication into the Request/Response model, i.e., turns all asynchronous Android IPC communication into Request/Response model. Support for each communication mechanism (e.g., Android IPC type) is enabled by the included \emph{IPCStack} implementations, allowing for new communication mechanisms to be added later through new implementations of \emph{IPCStack}. Thus, LCM can easily be extended later to support AF communication mechanisms other than existing Android IPC, although in this paper we focus only on improving existing Android IPC used for AF communication.

Upon receiving a Request, LCM selects a local communication method and the corresponding stack implementation for establishing connectivity. The selection of the most suitable communication method (from available \emph{IPCStack} implementations) can be based on any meaningful condition and realized as the selection algorithm. However, the current implementation simply selects the one set by the AF developer, i.e., the local communication mechanism set with highest priority (in a ordered list of supported mechanisms provided), by the developer of the requested AF.

%\begin{figure}[h]
%  \centering
%  \includegraphics[width=\linewidth]{commstack}
%  \caption{Code snippet showing the selection and utilization of corresponding local communication stack}
%  \Description{}
%  \label{fig:commstack}
%\end{figure}

\begin{figure}
  \centering
  \includegraphics[width=\linewidth]{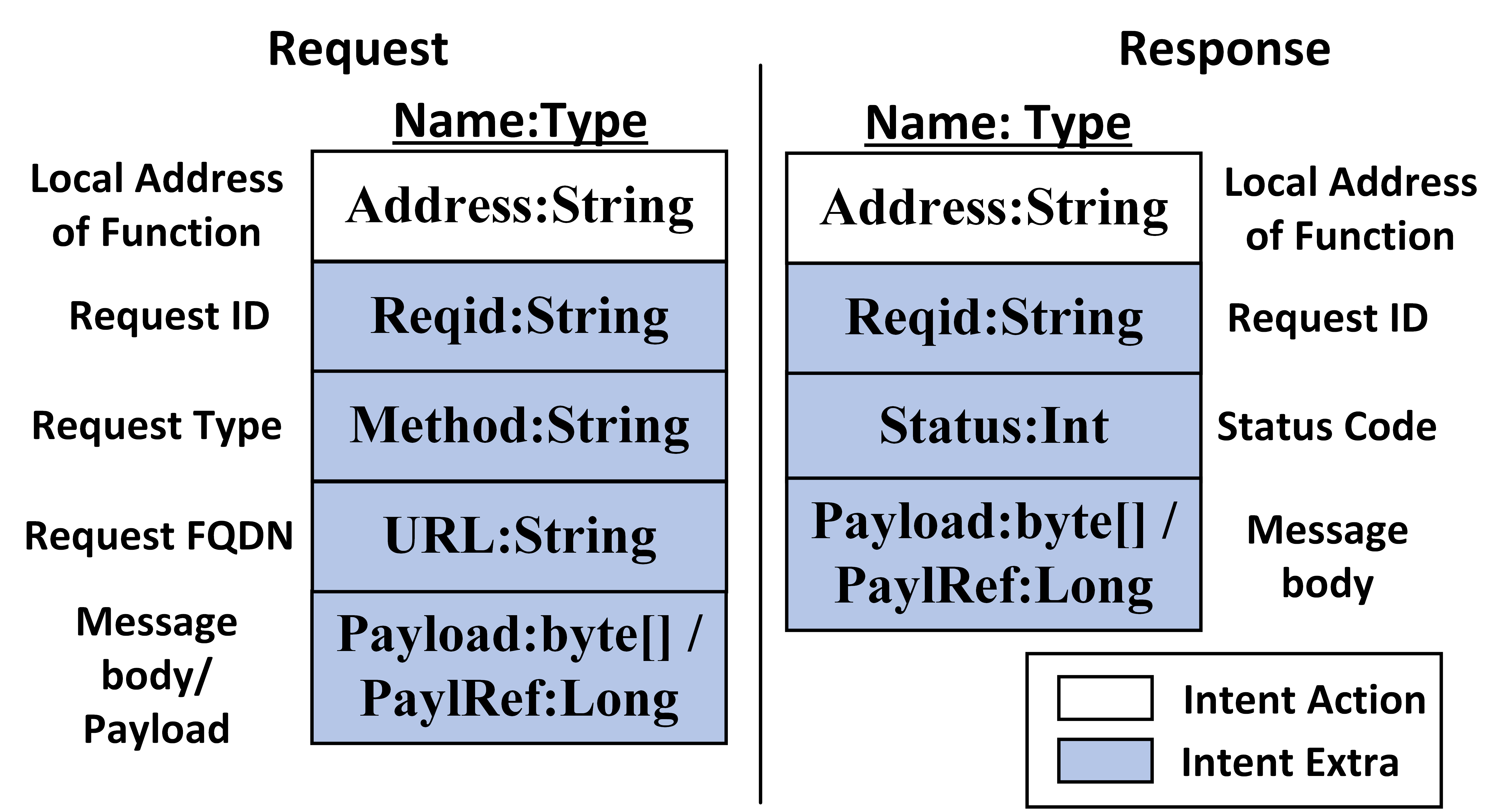}
  \caption{Message format - using Android Intent IPC}
  \Description{}
  \label{fig:ipcpacket}
\end{figure}

The most prominent abstraction to the underlying Binder IPC kernel driver is Android Intent \cite{10.1145/2664243.2664264}. LCM uses Android Intent \cite{AndroidDevelopersIntent} messages as the default unified intra-app (process local Intents \cite{AndroidDevLBroadcastManager} with improvements made through extensions to android Intent IPC introduced in Section \ref{sec:applayerheap}) and inter-app (Android global broadcasts \cite{AndroidDevBroadcasts}) communication mechanism due to its flexible, easy to use and optimized management procedures that are widely used throughout the Android OS \cite{10.1145/2664243.2664264}.  Figure \ref{fig:ipcpacket} shows the message format used when sending AF (request/response) messages over Android Intent IPC. Data in messages are stored as Key-Value pairs, with field names as keys and their values in the corresponding value fields. The Intent action field is used as the AF address field, while extra fields are used for storing all other fields, including the payload.

The '\emph{Reqid}' field contains an identifier that uniquely identifies the request (which is also in the response for matching responses with requests). The '\emph{Method}' and '\emph{URL}' fields contain the request type, which follows the web request type convention (e.g., GET, POST), and the complete AF resource/service URL (if in the Request), respectively. The '\emph{Payload}' field contains the body of the message. The '\emph{Status}' filed contains the status code of the response, following the same convention of the '\emph{status code}' values of HTTP \cite{httpstatuscodes}.

\subsection{Application-Layer Heap}\label{sec:applayerheap}

A major drawback of using any Android IPC (e.g., Intent, Binder, Content Provider) for communicating among AFs locally when not offloading, is the resulting significant increase in memory overhead when transferring large payloads \cite{10.1145/2436196.2436200} and inefficient memory management which eventually lead apps to crash. This renders Android IPC impractical to be used for some applications such as, continuous sensing and streaming apps, that require sending a large amount of data between AFs. 

Specifically, the focus of this section is on further improving communication efficiency of AFs within the same application (intra-app), for the application scenario of runtime bandwidth and latency requirements of intra-app AF communication are higher, and the messages sent between inter-app AFs do not include large payloads (decided by app developer at design-time - AFs with high bandwidth and latency requirements reside in the same Android app locally in the mobile device). Therefore, we continue to use Android IPC (Intent abstraction of Binder) in its original form for inter-app communication.

\begin{figure}
  \centering
  \includegraphics[width=0.8\linewidth]{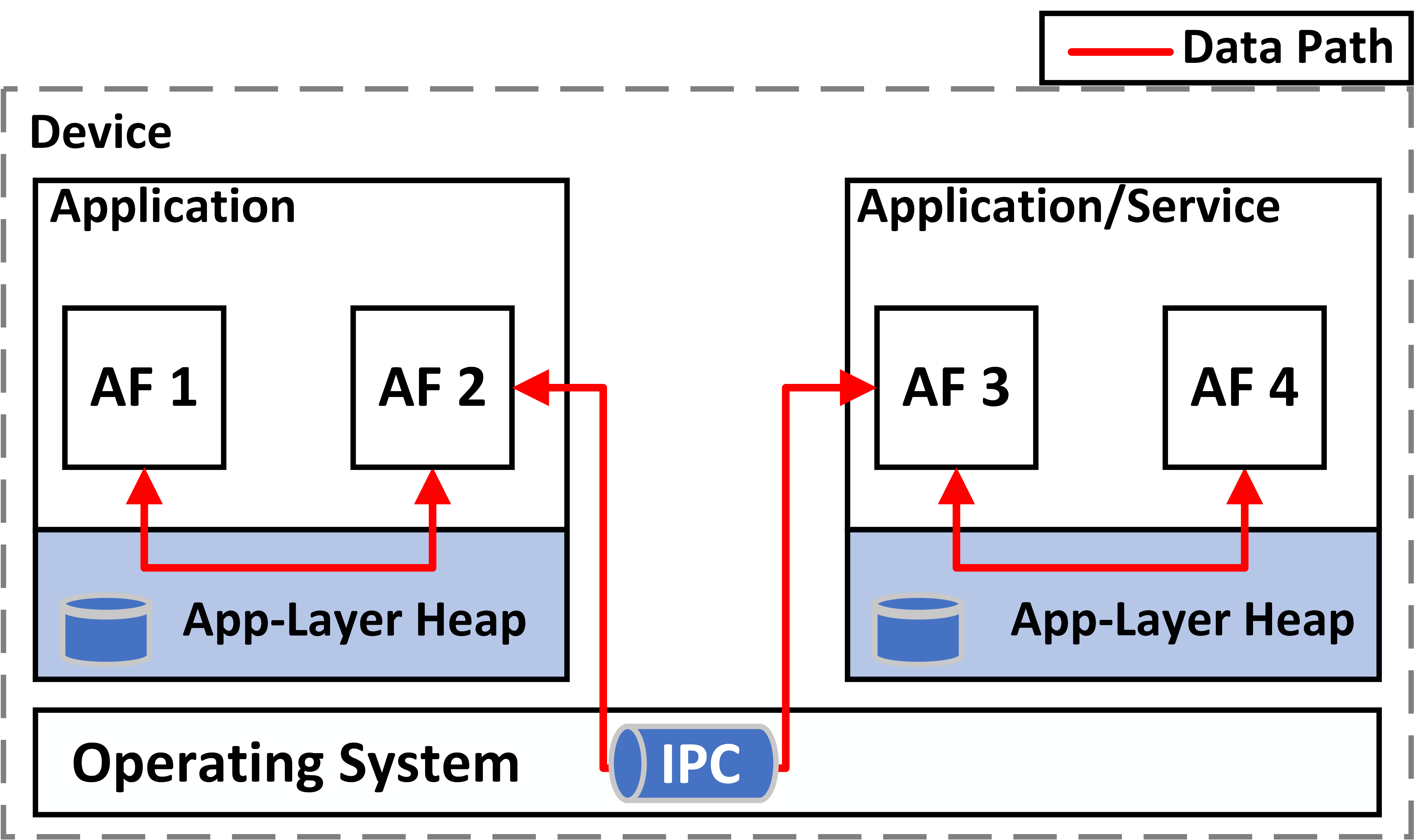}
  \caption{App-layer Heap overview}
  \Description{}
  \label{fig:heapoverview}
\end{figure}

We introduce an alternative intra-app AF-to-AF communication mechanism for overcoming aforementioned limitations imposed by Android IPC mechanisms, while still enabling flexible offloading of AFs into the network. Application-layer Heap introduced in Figure \ref{fig:heapoverview} provides a shared memory space for storing and communicating data between AFs. When AFs communicate within the same app, \emph{Payload} data stored on Application-layer Heap are not shared with Android IPC or the broader system and, does not leave the application process. Moreover, Application-layer heap maintains only one copy of data and provides REACT framework with complete control over the management of data (e.g., store, delete), i.e., it stores the payload when initializing a message and deletes it immediately after transmission is complete. This removes any unwanted copying and other inefficiencies when managing large payloads in memory by Android IPC. 

Alternatively, it is also possible for AFs to directly access the Application-layer Heap, such that, operations in memory (store, amend, delete) can be manually performed by AFs and, in turn provide/receive the reference to payload in memory to/from REACT for transmitting, respectively. However, details on how this can be done using tools provided by the REACT framework library is not disclosed here as it is out of scope of this paper. Therefore, in the rest of the paper, all memory operations are assumed to be performed automatically by the LCM. 

%\begin{figure}[h]
%  \centering
%  \includegraphics[width=\linewidth]{heapformat}
%  \caption{Format of Request/Response messages using App-layer Heap}
%  \Description{}
%  \label{fig:heapformat}
%\end{figure}

When communicating using Application-layer Heap, only a reference ('\emph{PaylRef}') to the (non-empty) payload is transmitted between AFs, as can be seen in the format of request/response messages in Figure \ref{fig:ipcpacket}. Given the improved performance when transmitting small messages \cite{10.1145/2436196.2436200}, REACT uses Android app-local Intent IPC \cite{AndroidDevLBroadcastManager} for communicating Application-layer Heap based messages that only contain references ('\emph{PaylRef}') in the message body.  Once, the message reaches its destination, the payload data is automatically retrieved from the Application-layer Heap and provided to the corresponding AF, before releasing the memory space occupied by the message payload. 

It is imperative that application data is localized and external exposure to them are limited (outside the app process) as much as possible, maintaining process-based isolation of data in Android \cite{10.1145/3307334.3326095}. Therefore, request/response payload data stored in Application-layer Heap is only accessible locally within the app, and the data itself does not leave the app at any point during intra-app AF communication, while the reference information included in the IPC messages is not useful outside the app. Widely used existing secure web protocols, such as HTTPS, may be used when offloading AFs to web services for securing application data against attacks. 

Any underlying Application-layer Heap implementation, 1) maintains only one copy of data in the device, 2) contains data within the app process, 3) provides LCM with means for storing and retrieving data, for sending and receiving requests and responses at the function borders, respectively. It is assumed that 3) provides a '\emph{PaylRef}' value per each stored payload data for uniquely identifying and accessing.

\begin{figure}
  \centering
  \includegraphics[scale=0.85]{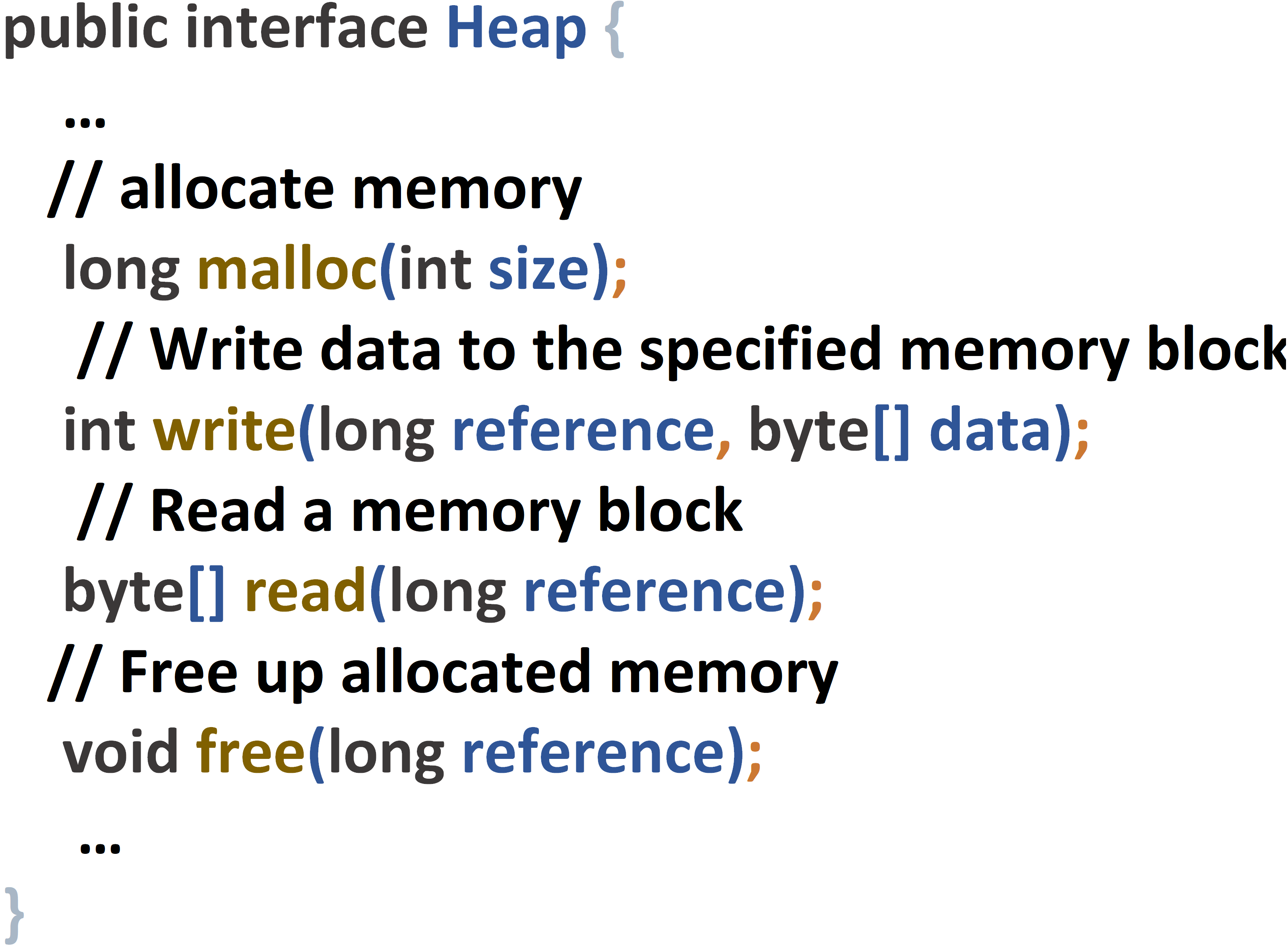}
  \caption{Code snippet of the Heap interface}
  \Description{}
  \label{fig:heapinterface}
\end{figure}

REACT library provides a common interface (Heap interface - code snippet shown in Figure \ref{fig:heapinterface}) which can be used for implementing and providing application-layer heap implementations to LCM, i.e., the interface can be used for providing new application-layer heap implementations, other than the one provided. Any object that implements this interface can be added to LCM using the methods provided by the library. The '\emph{malloc}' and '\emph{write}' methods are used for storing payload data, by allocating a block of space (that matches the size of the data) in memory and writing the data to it, respectively, before adding the reference number returned by '\emph{malloc}' to the '\emph{PaylRef}' field in the message body (Figure \ref{fig:ipcpacket}). Likewise, '\emph{read}' and '\emph{free}' method are used for retrieving data and freeing/deleting used memory blocks, respectively.

\emph{ByteArrayHeap} is a simple implementation of a Heap, provided by REACT library, that uses a Java one dimensional \emph{ByteArray} as the memory space. \emph{ByteArrayHeap} is set as the default Heap implementation of REACT runtime. Memory blocks are reserved within the continuous space of the \emph{ByteArray}, between the first byte and the last. The reference of a memory block is the index of its first byte within the \emph{ByteArray} (as allocated by \emph{malloc}), which therefore can be used to uniquely identify the block and its starting point in the larger array (as one byte can’t belong to more than one block at a given point in time - block spaces do not overlap). Likewise, a memory block's "end" border index can easily be calculated by simply adding the size of the payload data (in bytes) to its block start index (the \emph{reference}). In what follows, we present the algorithms being used for implementing the methods in Figure \ref{fig:heapinterface}, and other procedures needed for managing the \emph{ByteArray} Heap implementation. 

\begin{figure}
  \centering
  \includegraphics[width=0.8\linewidth]{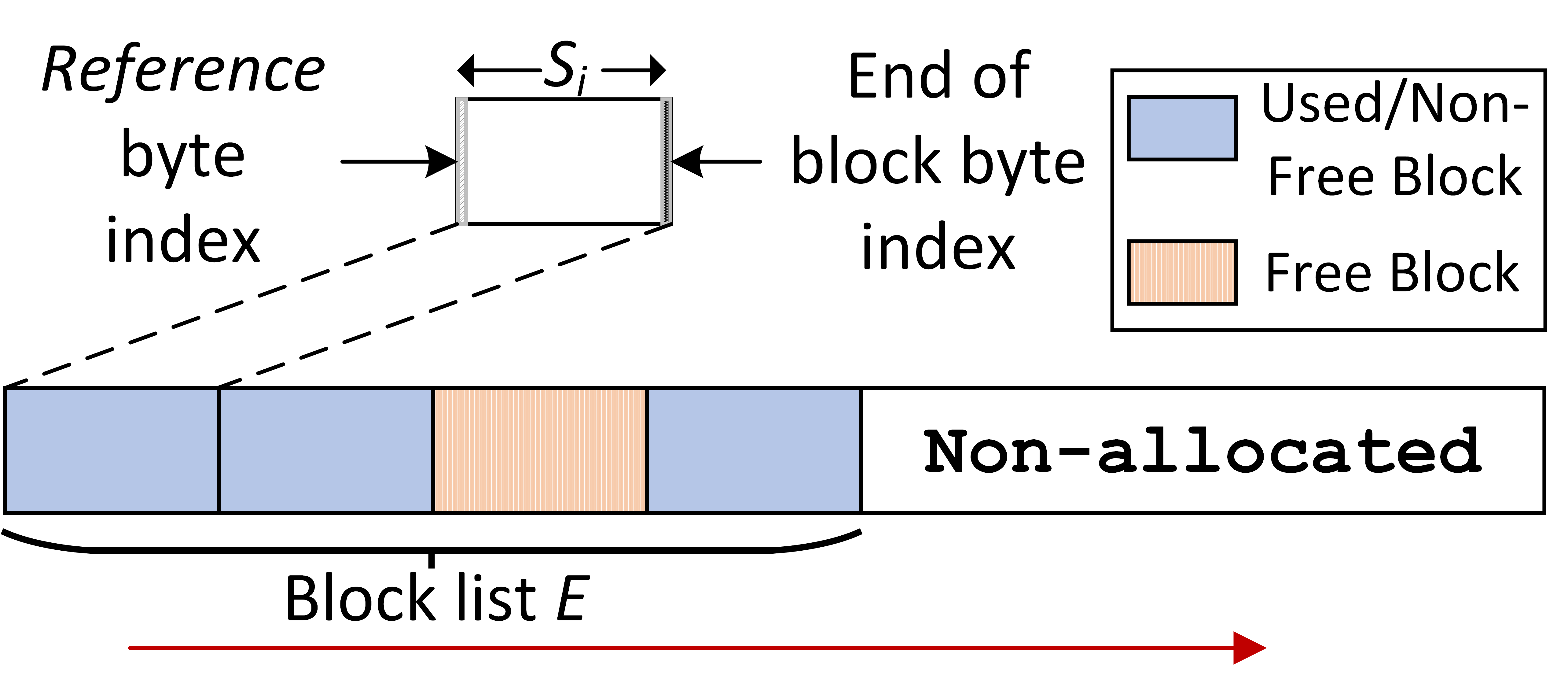}
  \caption{ByteArrayHeap block allocation overview}
  \Description{}
  \label{fig:memblocks}
\end{figure}

Portions of this continuous byte space is allocated through compartmentalizing the array into separate memory blocks and maintaining a record them, as depicted in Figure \ref{fig:memblocks}. For keeping track of the blocks and the order they are stored in (shown by arrow in Figure \ref{fig:memblocks}), a list $E$ (a \emph{LinkedList} \cite{oraclelinkedlist}) of memory \emph{Block} objects corresponding to all existing memory blocks is maintained. Each \emph{Block} element $i$ in the list stores its '\emph{reference}', the '\emph{size}' ($S_i$) of the block and its '\emph{status}', as member fields. The status indicates weather the block is currently being used ('non-free') or not ('free' - a freed portion of memory that is not being used). It is this list that is manipulated when managing the \emph{ByteArrayHeap}, as opposed to making changes directly to the underlying \emph{ByteArray} when performing aforementioned operations. The actual bytes in the \emph{ByteArray} never get erased, but instead get overwritten with new bytes, according to the new block structure, when the corresponding 'free' portion of memory gets assigned to a new reallocation. 

\emph{malloc} reuses 'free' blocks after resizing to match the new block size, releasing the remainder of bytes back to heap as 'free' memory, if any. However, if \emph{ByteArrayHeap} left unmanaged, over time, the operations described above can potentially leave the underlying \emph{ByteArray} increasingly fragmented, e.g., due to residue bytes of reused blocks left unused. Thus, the adjacent 'free' blocks gets periodically \emph{merge}d and 'free' blocks at the end of the list gets deleted (getting released to the non-allocated space). Finally, when retrieving payload data (with the \emph{read} method), only the bytes within the border (bytes from '\emph{reference}' location to end location of the block inclusive) of the corresponding block is read.

In summary, \emph{LCM} enables \emph{Execution Engine} to flexibly switch between IPC provided by the Android system and network requests, based on AF offloading decisions made by the \emph{Offload Decision Making Engine}. Moreover, application-layer heap enables, REACT localise intra-app communication data and, transport large payloads using Android IPC between intra-app AFs efficiently. 

\section{Implementation}
When building a REACT prototype, we have used the Android Volley HTTP library \cite{androidvolley} for creating and handling HTTP requests/responses that are sent between offloaded services and application functions that are running on the device. We have modified the Volley library for implementing the functionality of the \emph{Execution Engine} of the REACT architecture (in Figure \ref{fig:architecture}), incorporating interfaces to the \emph{Local Decision Making Engine} and the \emph{Local Communication Manager}, for indirecting requests to corresponding locally executing application functions, based on the offloading decisions made by the \emph{Offloading Decision Making Engine}. REACT uses \emph{RequestQueue} of the Volley library for queuing and processing inter-AF requests.

\textbf{Microservice/AF Addressing}: REACT identifies all application components, AFs/microservices running locally on the devices and web services in the network by their FQDN (Fully Qualified Domain Name). Locally executing AFs/Microservices may use the same FQDN (as their \emph{address}), as their matching services (that are used for offloading) in the network. Such an addressing scheme simplifies the identification and mapping of addresses between locally executing AFs and corresponding services in the network, providing a one-to-one mapping. However, following the Java/Android application naming convention \cite{androidappid}, addresses of all locally executing AFs use the reverse FQDN of its web service counterpart. For example, an AF may be addressed "\emph{com.example.myapp.process}", when the address of the web service counterpart is available on "\emph{process.myapp.example.com}". In this example, the \emph{application ID} is com.example.myapp and the process function is provided by the application provider ("\emph{com.example.myapp}"). Likewise, the developer may also include REACT compatible mobile AFs/microservices along with its web service counterparts provided by a third-party application provider (e.g., "com.example2.thirdpartyapp"), i.e., AFs using REACT wrapper class, distributed by the provider as a third-party android library. This allows REACT to discover locally running AFs and indirect corresponding incoming requests to corresponding functions.

\section{Evaluation}
We have implemented an REACT prototype for demonstrating how mobile applications can be developed as a collection of AFs (following the microservices architecture), and how they can be managed dynamically at runtime while adapting to contextual changes. Mainly, we focus the on the flexibility provided by the framework for offloading AFs and managing the communication between locally executing AFs and offloaded AFs, using the application scenario presented in Section \ref{sec:appusecase}. Moreover, we evaluate the performance of the communication mechanism introduced for inter-app AF communication. 

The results presented shows the flexibility in managing AFs running on a user's primary device, while adapting contextual changes. The REACT implementation used for our evaluation is based on Android v.10. We have used Google Pixel 2 XL, Nvidia Shield Pro devices, and a KVM VM deployed in an Openstack environment, which are all connected to the same 2.5GHz Wi-Fi access point.

\subsection{Application Use Case} \label{sec:appusecase}

\begin{figure}
  \centering
  \includegraphics[width=\linewidth]{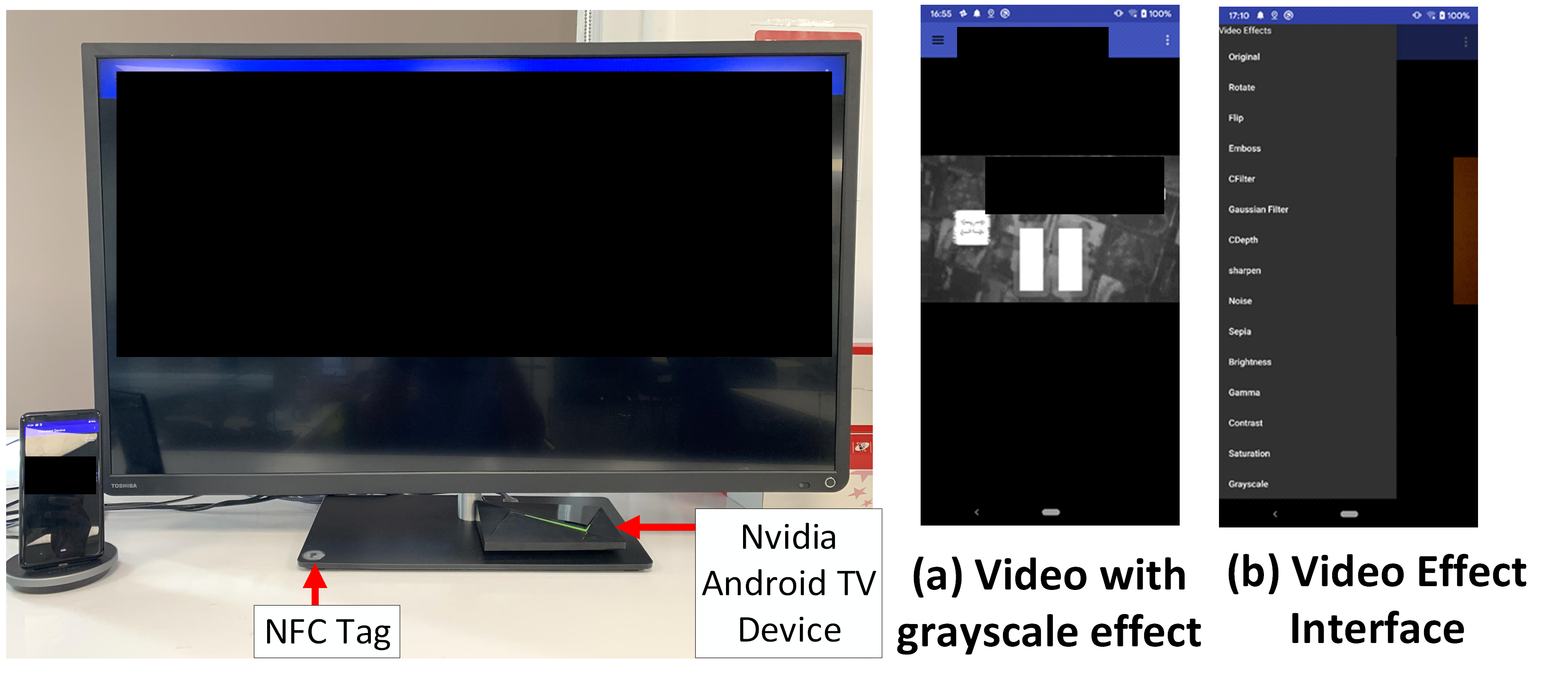}
  \caption{REACT enabled Video streaming app}
  \Description{}
  \label{fig:TDapp}
\end{figure}

%\footnote{Available on Android Play App Store}
%\footnote{App name and logos have been ommited for the blind review}
For demonstrating and evaluating the capabilities of the REACT framework, we have implemented an Android video viewing application (setup and app shown in Figure \ref{fig:TDapp}). Using the REACT API, the application has been developed as a collection of three AFs, namely, "\emph{Control}" AF, "\emph{Display}" AF and "\emph{Process}" AF. The \emph{Control} AF provides the user with a control interface, allowing the user to perform various video control actions (e.g., start/pause video). The "\emph{Process}" AF receives the video from a video source in the network, processes the video by applying an effect (selected by the user), and provides the output to the "\emph{Display}" AF, forming a service chain. The "\emph{Display}" AF displays the video on the screen. However, in this paper, we do not discuss the details of application-specific internal AF implementations themselves (e.g., video synchronization between AFs, video related control signalling between AFs), but instead focuses only on the effect REACT has on the mobile device during management and flexible execution of those AFs in a distributed environment. The video app is run on a Google Pixel 2 XL mobile device, and we fix the resolution of video to 360x640 for all experiments.

% DONE: Specify which web service in the network is being used
% DONE: specify the specification of the video being used
% We have implemented a second \emph{Process} function that runs on android, which provides video processing services over a web interface (using the NanoHTTPD \cite{elonen2012nanohttpd} android embedded HTTP server).
\textbf{Web services in the network}: We assume that a matching \emph{Process} function and a \emph{Display} function are available in the network (i.e., accessible through the WiFi network). A \emph{Process} service is implemented as a C++ application, and deployed in a Openstack environment which serves \emph{Process} service requests over the Apache web server \cite{serverapache}. Likewise, a matching Android \emph{Display} service/AF is deployed on the NVidia Android TV device that is connected to a TV screen, which displays the video on the connected TV, based on requests received from the \emph{Control} AF over the web interface (using the NanoHTTPD \cite{elonen2012nanohttpd} android embedded HTTP server). When the mobile device offloads to the \emph{Display} service, it retrieves the processed video frames from the \emph{Process} service deployed in the network, leaving only the \emph{Control} AF on the mobile device.

\textbf{Offload Decision Making}: We have employed a simple offloading policy that takes the network connectivity and NFC tag readings as input. The \emph{Process} AF is offloaded only when the user is connected to the home network. Likewise, the \emph{Display} AF is offloaded (and pulled back to mobile device) only on reading a known NFC tag associated with a display device with \emph{Display} service installed, while at the same time connected to the HOME Wi-Fi network. Both policies are manually programmed into the \emph{Offload Decision Making Engine}.

\subsection{Power Consumption}
In order to quantitatively demonstrate the overall performance of the REACT framework and the adaptability of the application that use the framework, we analyse the changes in instantaneous power consumption. For measuring the power consumption of the Android device, we use the Android's \emph{BatteryManager} \cite{10.1007/978-3-319-46523-4_18} \cite{androidbatterymanager}.

\begin{figure}
  \centering
  \includegraphics[width=\linewidth]{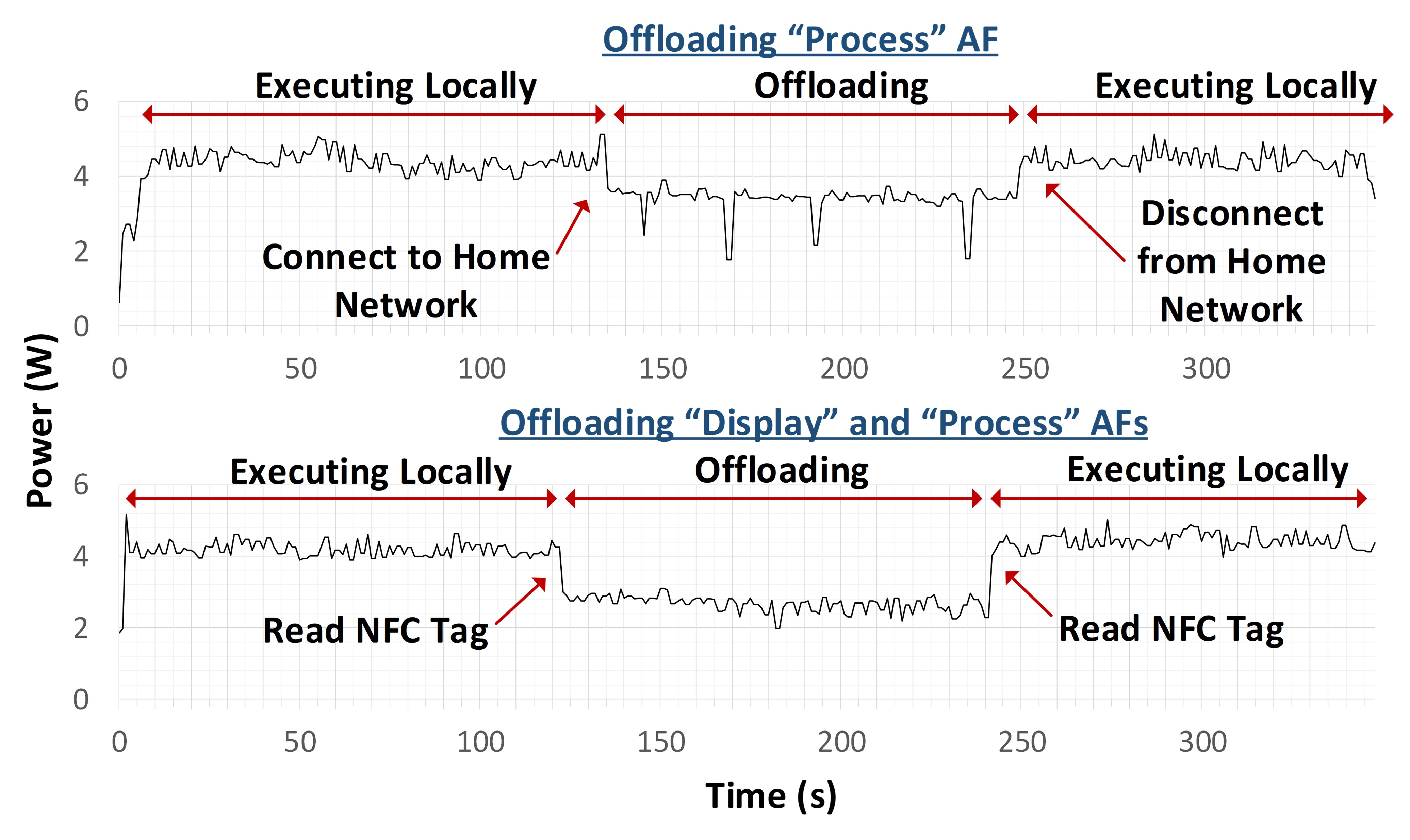}
  \caption{Power usage when executing \emph{Process} AF (with "Sharpen" effect enabled), local vs offloading}
  \Description{}
  \label{fig:sharpenpowertime}
\end{figure}

Figure \ref{fig:sharpenpowertime} shows the changes in power consumption as a result of dynamic AF offloading, triggered by the changes in network connectivity and NFC readings. As the user connects to the home Wi-Fi network, REACT offloads the \emph{Process} AF, as instructed by the \emph{Offloading Decision Making Engine}, reducing the instantaneous overall app power consumption by $\sim$1.01 watts on average. Then, as the user disconnects from the home Wi-Fi network, the \emph{Process} function is dynamically initiated locally and the communication between the local \emph{Process} and \emph{Display} functions are re-established automatically as the \emph{Process} AF registers against the function catalogue as a locally available AF. Likewise, when offloading \emph{Display} AF (along with \emph{Process} AF), based on NFC tag readings, reducing the instantaneous overall app power consumption by $\sim$1.53 watts on average. 

\begin{figure}
  \centering
  \includegraphics[width=0.8\linewidth]{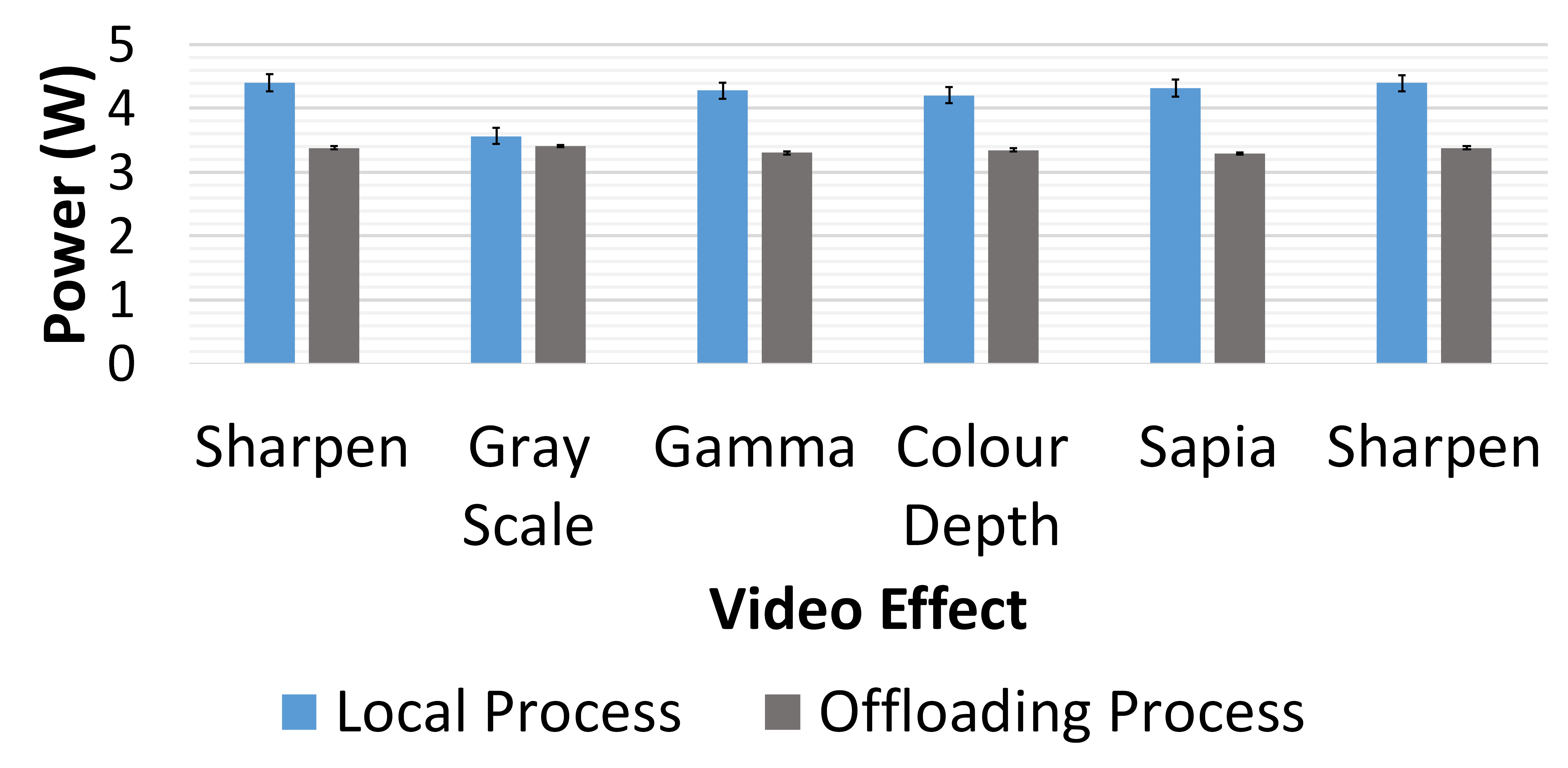}
  \caption{Average power consumption}
  \Description{}
  \label{fig:powervseffect}
\end{figure}

Improvements that may be made in power usage by offloading, depends on the resource (and communication) intensity of the AF being offloaded. However, choosing or developing suitable AFs towards improving power consumption is out of the scope of this paper, and relies solely on the static polices programmed in \emph{Offload Decision Making Engine} for showing REACT's dynamic and flexible management of independent AFs. Figure \ref{fig:powervseffect} shows the average power consumption (per second) of the video viewing application running on the mobile device, depending on the video effect (with varying computing intensity) that's being used. Specifically, on average, offloading can save power when using the "Sharpen" effect, and less beneficial to offload when adding the "Gray scale" effect.

\subsection{Memory Overhead}
% DONE: Find out what Runtime measures. Is it app usage or JVM/Dalvik VM usage?
We used \emph{Runtime} \cite{androidruntime} on Android to measure the Random Access Memory (RAM) usage within the application memory space, and used \emph{MemoryInfo} \cite{androidmeminfo} for measuring the system-wide RAM usage.

\begin{figure}
\centering  
\subfigure[Memory usage without app-layer heap - from app level]{\label{fig:appmem_withoutheap}\includegraphics[width=40mm, height=30mm]{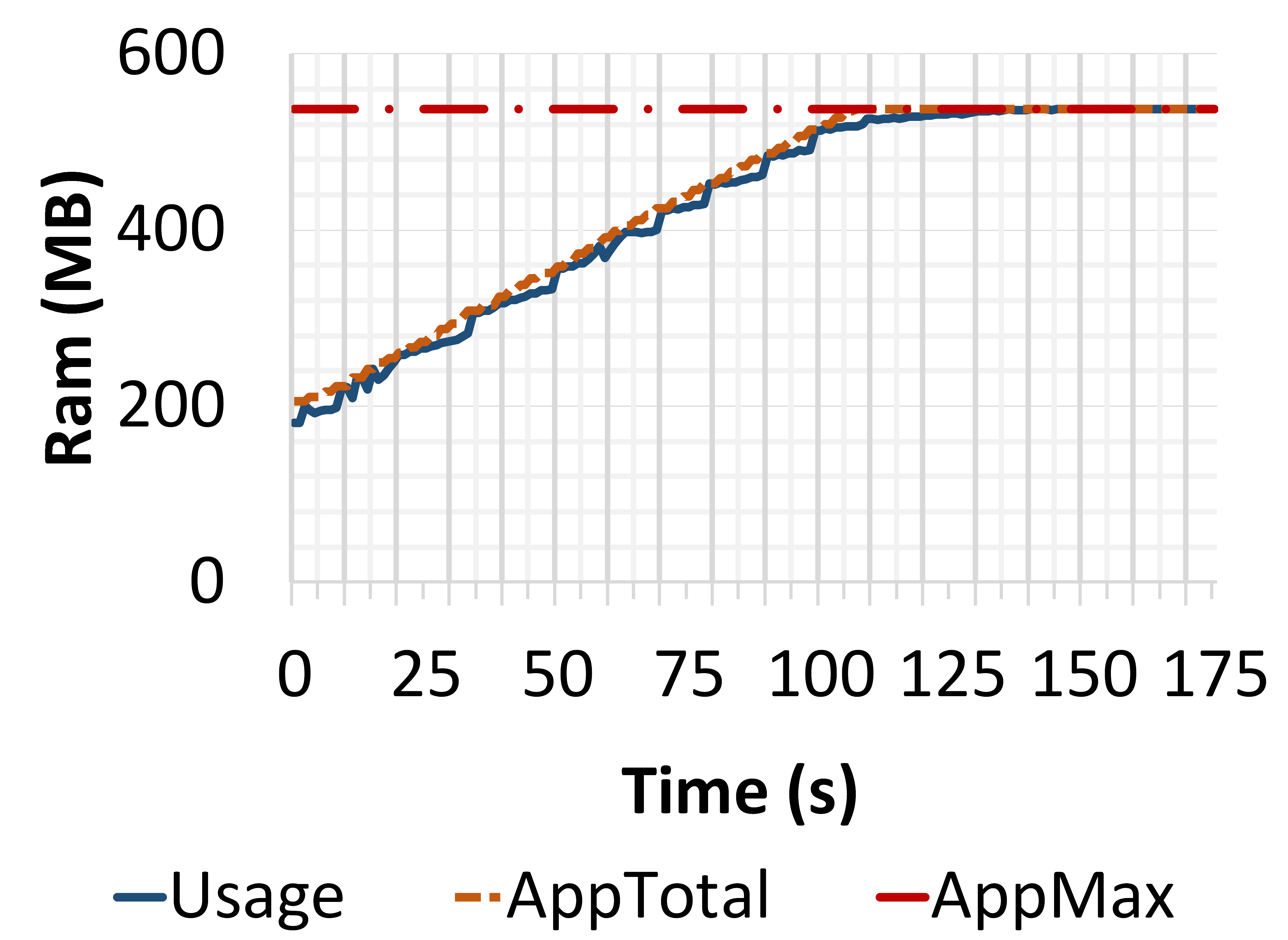}}
\subfigure[Memory usage without app-layer heap - from system level]{\label{fig:sysmem_withoutheap}\includegraphics[width=40mm, height=30mm]{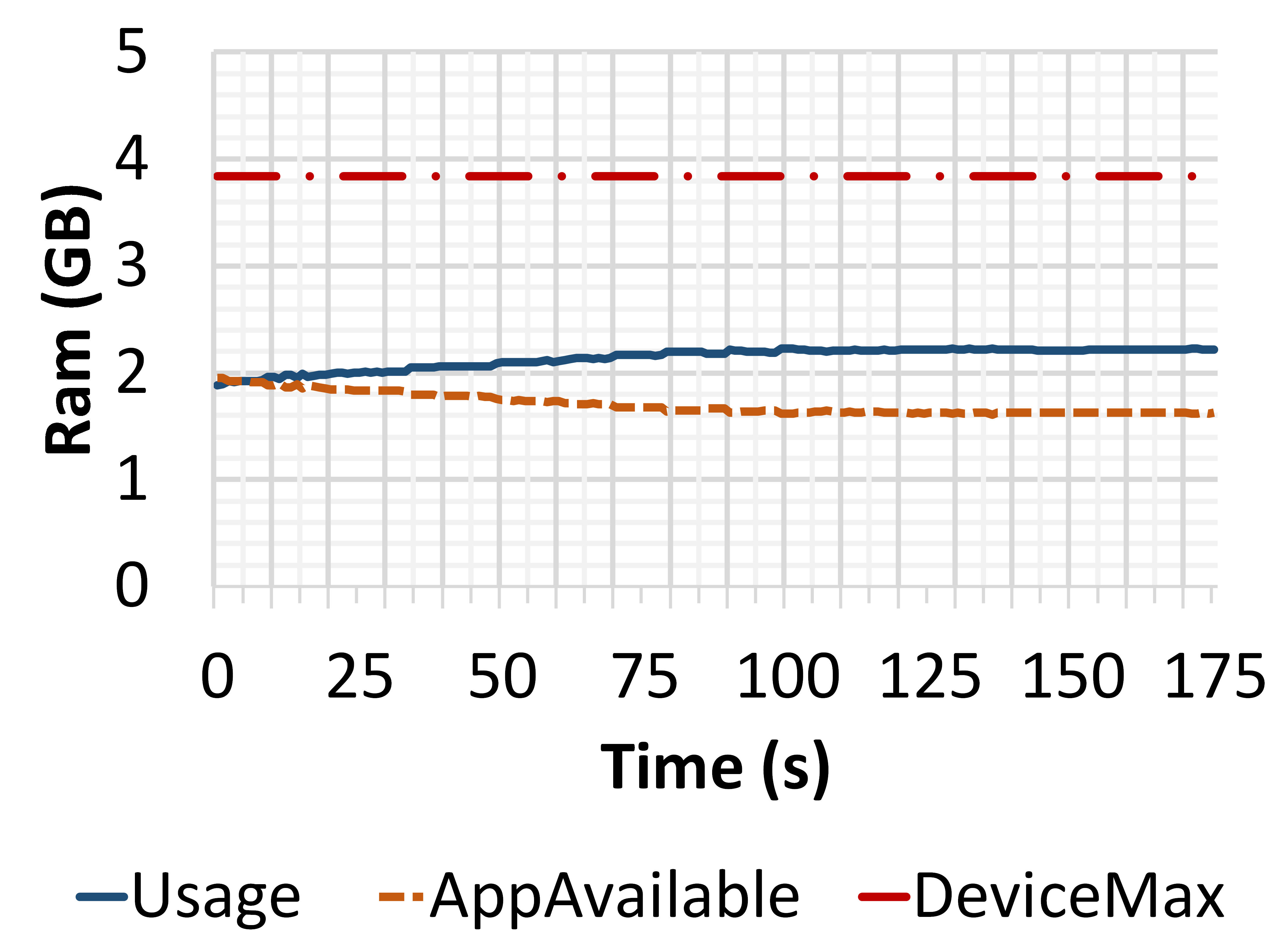}}

\subfigure[Memory usage with app-layer heap - from app level]{\label{fig:appmem_withtheap}\includegraphics[width=40mm, height=30mm]{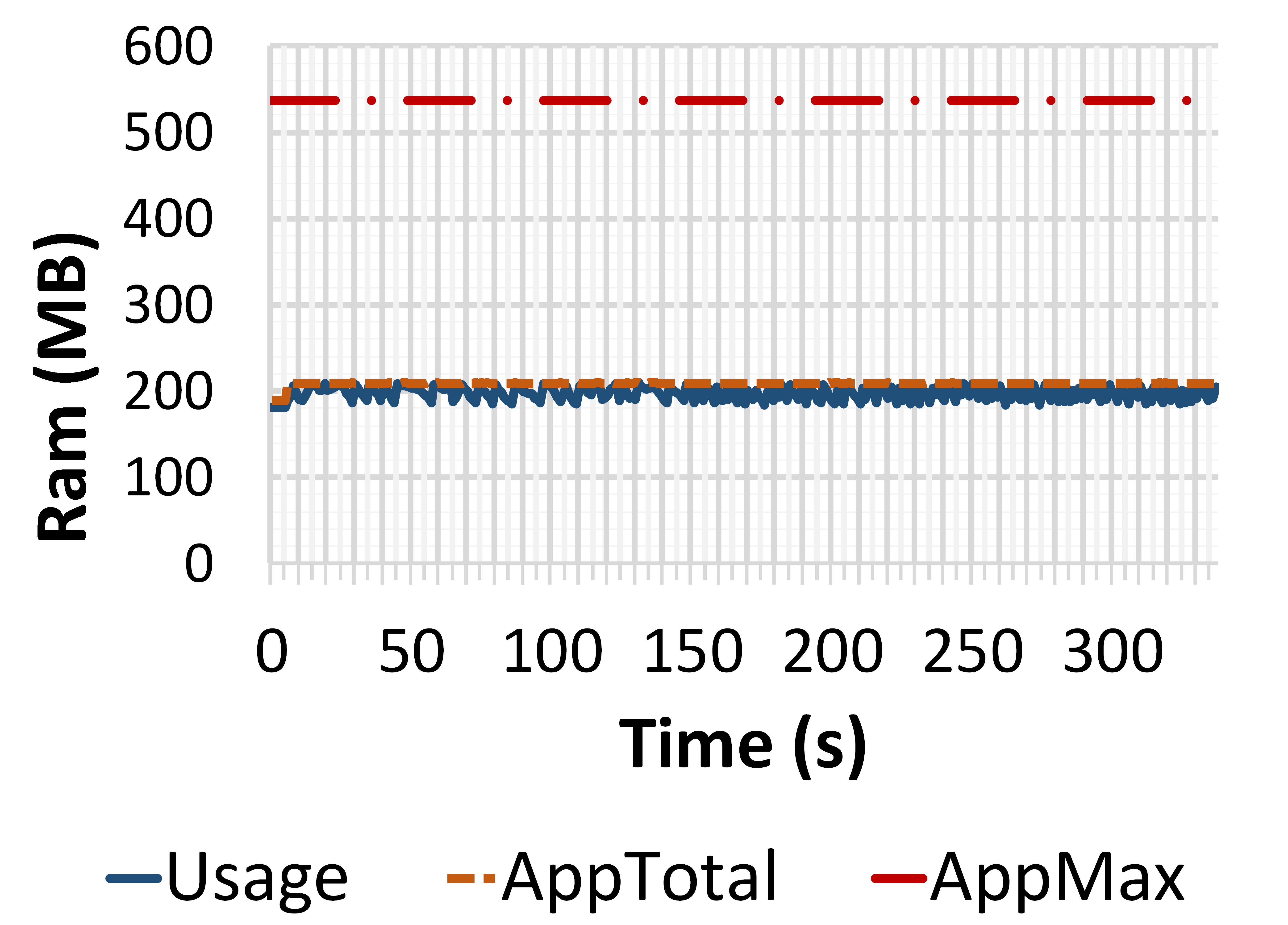}}
\subfigure[Memory usage with app-layer heap - from system level]{\label{fig:sysmem_withtheap}\includegraphics[width=40mm, height=30mm]{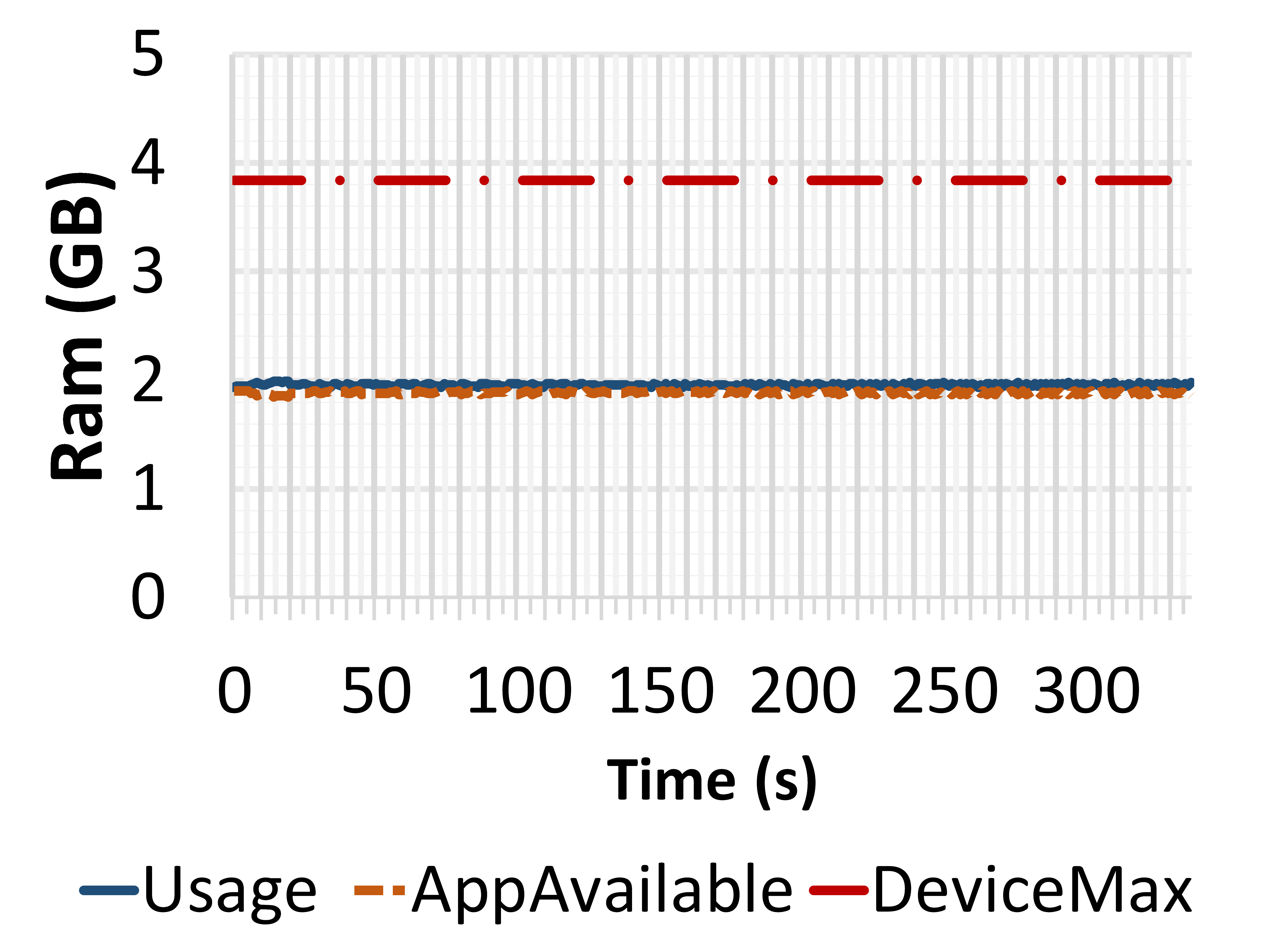}}
\caption{Memory overhead of the application from the application and system levels}
\end{figure}

%\begin{figure}[h]
%\begin{subfigure}[b]{0.2\textwidth}
%    \centering
%	% include first image
%  \includegraphics[width=\linewidth]{appram_noheap}  
%  \caption{App memory without app-layer heap}
%  \label{fig:appmem_withoutheap}
%\end{subfigure}%
%~
%\begin{subfigure}[b]{0.2\textwidth}
%  \centering
%  % include second image
%  \includegraphics[width=\linewidth]{appram_withheap}  
%  \caption{App memory with app-layer heap}
%  \label{fig:appmem_withtheap}
%\end{subfigure}%
%
%~
%
%\begin{subfigure}[b]{0.2\textwidth}
%  \centering
%  % include third image
%  \includegraphics[width=\linewidth]{sysram_noheap}  
%  \caption{System memory without app-layer heap}
%  \label{fig:sysmem_withoutheap}
%\end{subfigure}%
%~
%\begin{subfigure}[b]{0.2\textwidth}
%  \centering
%  % include fourth image
%  \includegraphics[width=\linewidth]{sysram_withheap}  
%  \caption{System memory with app-layer heap}
%  \label{fig:sysmem_withtheap}
%\end{subfigure}%
%\caption{Memory overhead at the application and system levels, with and without application-layer heap}
%\label{fig:fig}
%\end{figure}

%%%%% ----------------------------------------------------------------------
%\begin{figure}%
%    \centering
%    \subfloat[label 1]{{\includegraphics[width=3.82cm]{powervseffect} }}%
%    \qquad
%    \subfloat[label 2]{{\includegraphics[width=3.82cm]{powervseffect} }}%
%    \caption{2 Figures side by side}%
%    \label{fig:example}%
%\end{figure}

%%%%%%%-0-----------------------------------------------------------------

% DONE: the Ram Total and MAx labels in the graphs are confusing - use the same terms throughout the graphs
As shown by previous studies \cite{10.1145/2436196.2436200} (and as mentioned in Section \ref{sec:applayerheap}) Android IPC mechanisms (in our case Android Broadcast Intents) infer significant performance penalties when used for transferring large payload data. Specifically, this can be observed over time in the application memory usage, both at application level (shown in Figure \ref{fig:appmem_withoutheap}) and system level (shown in Figure \ref{fig:sysmem_withoutheap}). In both cases, memory consumption increases continuously over time, until the application crashes, after reaching the maximum amount of memory available to the application, $\sim$110 seconds after running the application.

\begin{figure}
\centering  
\subfigure[Avg. app-layer heap memory usage per second]{\label{fig:heapmemVSvideffect}\includegraphics[width=40mm, height=30mm]{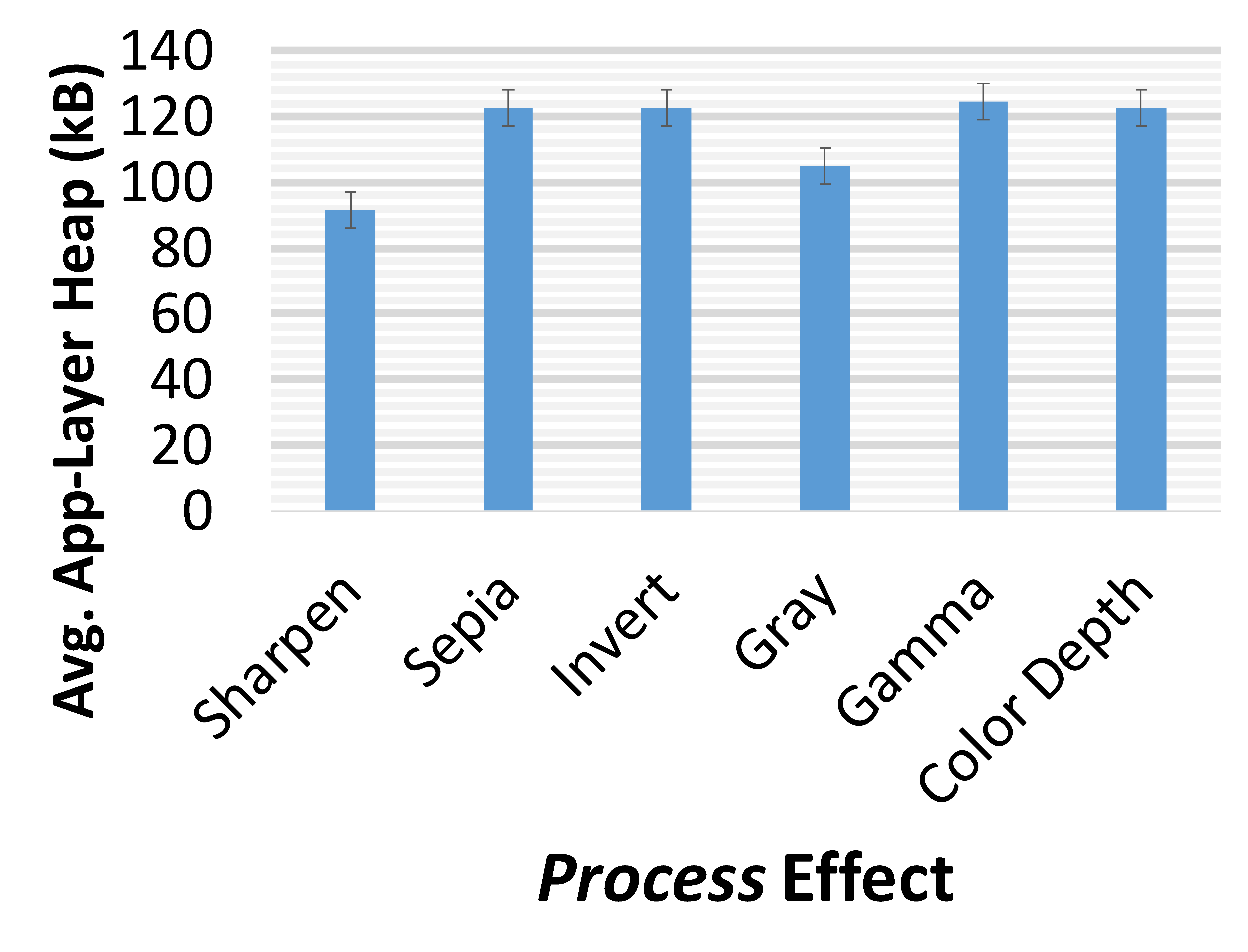}}
\subfigure[Number of blocks, with and without \emph{merge} enabled]{\label{fig:merge_withvswithout}\includegraphics[width=40mm, height=30mm]{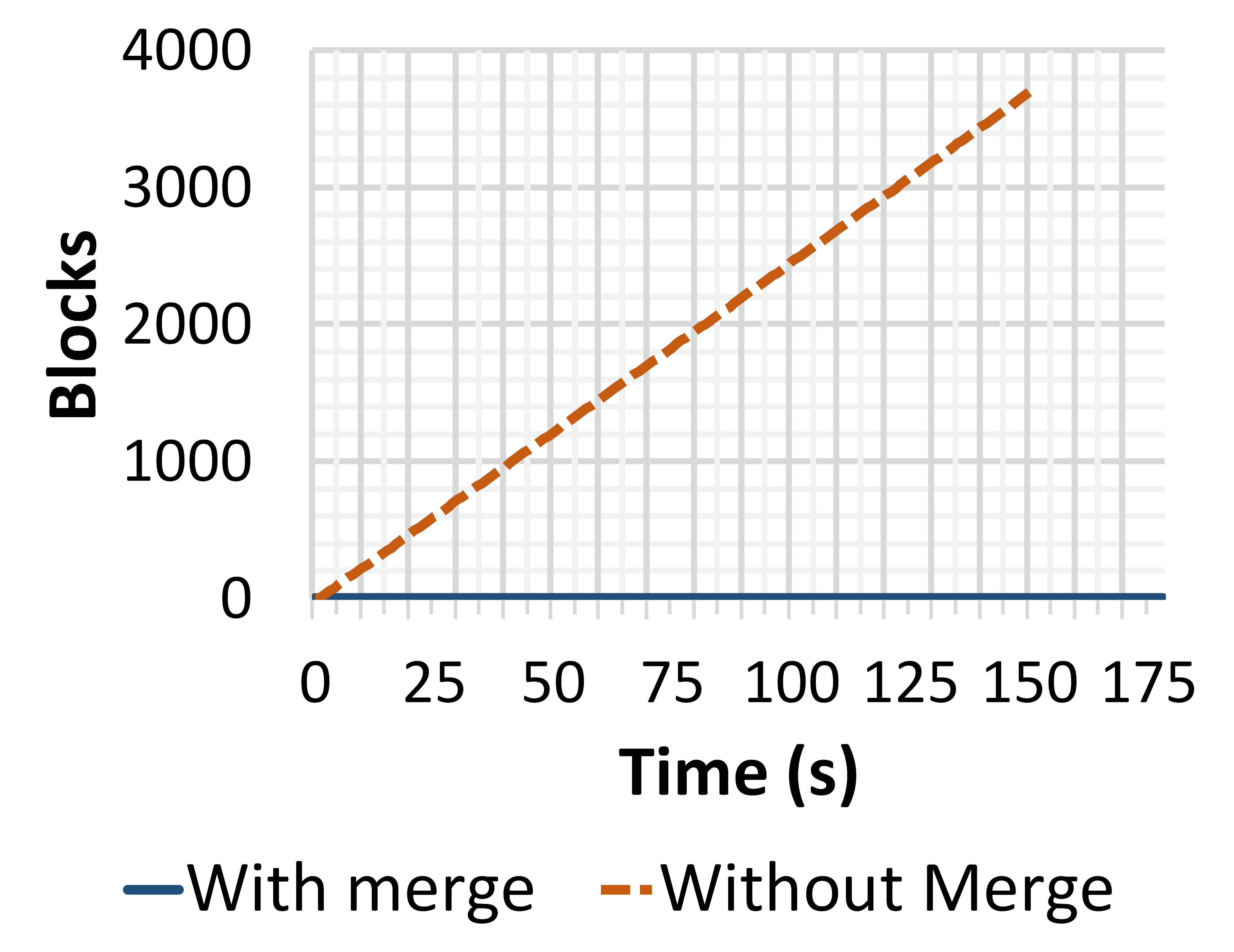}}
\caption{App-layer Heap memory usage}
\end{figure}

However, Figure \ref{fig:appmem_withtheap} and Figure \ref{fig:sysmem_withtheap} shows how memory usage at the application level and system level is being stabilized (at $\sim$200Mb) through our application-layer heap enabled AF communication mechanism, respectively. The shared memory space provided by the app-layer heap has enabled REACT to localise the application specific data in intra-app AF communication while also improving the efficiency of the Android IPC mechanism. The \emph{merge} procedures introduced in Section \ref{sec:applayerheap} keeps the application-layer heap defragmented. Figure \ref{fig:heapmemVSvideffect} shows the average app-layer heap usage per second. Figure \ref{fig:merge_withvswithout} shows how the number of created memory blocks increase when \emph{merge} is disabled, while enabling \emph{merge} procedures keeps the number of blocks to a minimum. As the number of memory blocks increase, the application eventually crashes ($\sim$150 seconds after running the application). %does this also throw OutOfMemoryError?

%In our video streaming app, the specific amount of memory being used depends on the size and resolution of the video, and the size of the video frames after adding an effect in the \emph{Process} AF.
%
%\begin{figure}[h]
%
%\begin{subfigure}{.1\textwidth}
%  \centering
%  % include fourth image
%  \includegraphics[width=0.5\linewidth]{heapmemVSvideffect}  
%  \caption{App-layer Heap memory usage}
%  \label{fig:heapmemVSvideffect}
%\end{subfigure}%
%
%~
%
%\begin{subfigure}{.1\textwidth}
%  \centering
%  \includegraphics[width=0.5\linewidth]{merge_withvswithout}
%  \caption{Number of blocks, with and without \emph{merge} enabled}
%  \Description{}
%  \label{fig:merge_withvswithout}
%\end{subfigure}
%
%\caption{App-layer Heap memory usage, with and without \emph{merge}}
%\label{fig:fig}
%\end{figure}

\section{Conclusion}

We have presented REACT, an Android-based application framework that enables mobile applications to be developed as a collection of loosely couple AFs/microservices (goal 1). REACT manages  microservices individually, allowing fine-grained application components to be dynamically offloaded to be executed by web services in the network (goal 3) based on contextual changes (goal 2). The \emph{Local Communication Manager} transfers requests and responses between microservices efficiently using our newly introduced application-layer heap enabled Android IPC mechanism, and enables seamless switching between network requests and Android IPC based requests transparently to the communicating microservices. Thus, the communication medium of microservice requests can be dynamically changed, depending on the locality of communicating microservices (goal 4). Our prototype implementation has proved that REACT enables the development of highly flexible microservices-based mobile applications that are highly dynamic and adaptable to contextual changes. The newly introduced, application-layer heap enabled Android IPC mechanism has proven to improve the efficiency of Android IPC, reducing memory overhead when communicating messages with large payloads between locally executing microservices. We expect REACT, distributed as an easy to use Android Library, to kick start the development of creative and useful microservices based Android mobile applications.

%% The acknowledgments section is defined using the "acks" environment
%% (and NOT an unnumbered section). This ensures the proper
%% identification of the section in the article metadata, and the
%% consistent spelling of the heading.
\begin{acks}
The authors would like to thank Dr. Dirk Trossen for the initial technical contribution to this work as team lead at InterDigital until December 2019.
\end{acks}

%%
%% The next two lines define the bibliography style to be used, and
%% the bibliography file.
\bibliographystyle{ACM-Reference-Format}
\bibliography{sample-base}

%%
%% If your work has an appendix, this is the place to put it.
\appendix

\end{document}